\PassOptionsToPackage{unicode}{hyperref}
\PassOptionsToPackage{hyphens}{url}
\PassOptionsToPackage{dvipsnames,svgnames,x11names}{xcolor}
\documentclass[12pt]{article}

\usepackage{amsmath,amssymb}
\usepackage{iftex}
\ifPDFTeX
  \usepackage[T1]{fontenc}
  \usepackage[utf8]{inputenc}
  \usepackage{textcomp} 
\else 
  \usepackage{unicode-math}
  \defaultfontfeatures{Scale=MatchLowercase}
  \defaultfontfeatures[\rmfamily]{Ligatures=TeX,Scale=1}
\fi
\usepackage{lmodern}
\ifPDFTeX\else  
\fi
\IfFileExists{upquote.sty}{\usepackage{upquote}}{}
\IfFileExists{microtype.sty}{
  \usepackage[]{microtype}
  \UseMicrotypeSet[protrusion]{basicmath} 
}{}
\makeatletter
\@ifundefined{KOMAClassName}{
  \IfFileExists{parskip.sty}{%
    \usepackage{parskip}
  }{
    \setlength{\parindent}{0pt}
    \setlength{\parskip}{6pt plus 2pt minus 1pt}}
}{
  \KOMAoptions{parskip=half}}
\makeatother
\usepackage{xcolor}
\makeatletter
\ifx\paragraph\undefined\else
  \let\oldparagraph\paragraph
  \renewcommand{\paragraph}{
    \@ifstar
      \xxxParagraphStar
      \xxxParagraphNoStar
  }
  \newcommand{\xxxParagraphStar}[1]{\oldparagraph*{#1}\mbox{}}
  \newcommand{\xxxParagraphNoStar}[1]{\oldparagraph{#1}\mbox{}}
\fi
\ifx\subparagraph\undefined\else
  \let\oldsubparagraph\subparagraph
  \renewcommand{\subparagraph}{
    \@ifstar
      \xxxSubParagraphStar
      \xxxSubParagraphNoStar
  }
  \newcommand{\xxxSubParagraphStar}[1]{\oldsubparagraph*{#1}\mbox{}}
  \newcommand{\xxxSubParagraphNoStar}[1]{\oldsubparagraph{#1}\mbox{}}
\fi
\makeatother

\usepackage{longtable,booktabs,array}
\usepackage{calc} 
\usepackage{etoolbox}
\makeatletter
\patchcmd\longtable{\par}{\if@noskipsec\mbox{}\fi\par}{}{}
\makeatother
\IfFileExists{footnotehyper.sty}{\usepackage{footnotehyper}}{\usepackage{footnote}}
\makesavenoteenv{longtable}
\usepackage{graphicx}
\makeatletter
\def\maxwidth{\ifdim\Gin@nat@width>\linewidth\linewidth\else\Gin@nat@width\fi}
\def\maxheight{\ifdim\Gin@nat@height>\textheight\textheight\else\Gin@nat@height\fi}
\makeatother
\setkeys{Gin}{width=\maxwidth,height=\maxheight,keepaspectratio}
\makeatletter
\def\fps@figure{htbp}
\makeatother

\makeatletter
\@ifpackageloaded{caption}{}{\usepackage{caption}}
\AtBeginDocument{%
\ifdefined\contentsname
  \renewcommand*\contentsname{Table of contents}
\else
  \newcommand\contentsname{Table of contents}
\fi
\ifdefined\listfigurename
  \renewcommand*\listfigurename{List of Figures}
\else
  \newcommand\listfigurename{List of Figures}
\fi
\ifdefined\listtablename
  \renewcommand*\listtablename{List of Tables} 
\else
  \newcommand\listtablename{List of Tables}
\fi
\ifdefined\figurename
  \renewcommand*\figurename{Figure}
\else
  \newcommand\figurename{Figure}
\fi
\ifdefined\tablename
  \renewcommand*\tablename{Table}
\else
  \newcommand\tablename{Table}
\fi
}
\@ifpackageloaded{float}{}{\usepackage{float}}
\floatstyle{ruled}
\@ifundefined{c@chapter}{\newfloat{codelisting}{h}{lop}}{\newfloat{codelisting}{h}{lop}[chapter]}
\floatname{codelisting}{Listing}

\makeatother
\makeatletter
\@ifpackageloaded{caption}{}{\usepackage{caption}}
\@ifpackageloaded{subcaption}{}{\usepackage{subcaption}}
\makeatother

\ifLuaTeX
  \usepackage{selnolig}  
\fi
\usepackage[]{natbib}
\usepackage{bookmark}

\IfFileExists{xurl.sty}{\usepackage{xurl}}{} 
\hypersetup{
  pdftitle={Unbiased Treatment Effect Estimation under Network Interference via Neighborhood-Excluded Cross-Fitting},
  pdfauthor={Haoyang Yu, Anqi Zhao, and Hanzhong Liu},
  pdfkeywords={Covariate adjustment, Design-based inference, High-dimensional covariates, Machine learning, Sample splitting},
  colorlinks=true,
  linkcolor={blue},
  filecolor={Maroon},
  citecolor={Blue},
  urlcolor={Blue},
  pdfcreator={LaTeX via pandoc}}

\newcommand{\anon}{1}

\usepackage{amsmath}
\usepackage{graphicx}
\usepackage{enumerate}
\usepackage{natbib}
\usepackage{multibib}
\usepackage{silence}
\usepackage{bm}
\usepackage{bbm}
\usepackage{url} 
\usepackage{amsthm}
\usepackage{amsfonts}
\usepackage{amssymb}
\usepackage{multirow}
\usepackage{booktabs}
\usepackage[justification=centering]{caption}
\usepackage{threeparttable}
\usepackage{graphicx}
\usepackage{float}
\usepackage{comment}
\usepackage{subcaption}
\usepackage[sort]{cleveref}
\usepackage{xr}
\usepackage{threeparttable}
\usepackage{xcolor}
\usepackage{colortbl}
\usepackage{titling}
\usepackage{algorithmic}
\usepackage{algorithm}
\usepackage{enumerate}

\usepackage{tikz}
\usetikzlibrary{calc}
\usetikzlibrary{arrows.meta,positioning,fit,shapes.misc}

\newcites{app}{References}
\bibliographystyleapp{agsm}

\theoremstyle{definition}

\newtheorem{assumption}{Assumption}
\Crefname{assumption}{Assumption}{Assumptions}
\newtheorem{theorem}{Theorem}

\newtheorem{definition}{Definition}

\newtheorem{corollary}{Corollary}
\newtheorem{remark}{Remark}
\newtheorem{example}{Example}

\newcommand{\mosik}{{\mathcal O_{*,\mik}}}

\newcommand{\iid}{\textup{i.i.d.}}\newcommand{\simiid}{\overset{\iid}{\sim}}\newcommand{\tyi}{\tilde Y_i(\bb)}
\newcommand{\tby}{\tilde {\bm Y}(\bb)}
\newcommand{\bb}{\bm \beta}
\newcommand{\hbdkci}{\hat{\bb}_{\dir,[k]}^\textup{CI}}
\newcommand{\hbikci}{\hat{\bb}_{\ind,[k]}^\textup{CI}}
\newcommand{\hbskci}{\hat{\bb}_{*,[k]}^\textup{CI}}

\newcommand{\hbsk}{\hat{\bb}_{*,[k]}}
\newcommand{\hbskv}{\hbsk^\textup{V}}
\newcommand{\hbikv}{\hat{\bb}_{\ind,[k]}^\textup{V}}
\newcommand{\hbdkv}{\hat{\bb}_{\dir,[k]}^\textup{V}}
\newcommand{\bsoraci}{\bb_*^{\ora\textup{-CI}}}
\newcommand{\bioraci}{\bb_\ind^{\ora\textup{-CI}}}
\newcommand{\bdoraci}{\bb_\dir^{\ora\textup{-CI}}}
\newcommand{\bsorav}{\bb_*^{\ora\textup{-V}}}
\newcommand{\biorav}{\bb_\ind^{\ora\textup{-V}}}
\newcommand{\bdorav}{\bb_\dir^{\ora\textup{-V}}}

\newcommand{\hfsk}{\hat f_{*,[k]}}

\newcommand{\mmtsk}{\mm_{\mtsk}}
\newcommand{\mg}{\mathcal G}

\newcommand{\thm}{Theorem}
\newcommand{\odi}{\omega_{\dir,i}}
\newcommand{\oii}{\omega_{\ind,i}}
\newcommand{\ogi}{\omega_{\gate,i}}
\newcommand{\mik}{\mathcal I_k}
\newcommand{\mikc}{\mathcal I_k^c}
\newcommand{\mtsk}{\mathcal T_{*,k}}
\newcommand{\lr}{\mathrm{LR}}
\newcommand{\hti}{\hat\tau_{\ind}}
\newcommand{\maxi}{\max_{1\leq i\leq n}}

\newcommand{\mm}{\mathcal M}
\newcommand{\mmi}{{\mm_i}}

\newcommand{\meani}{n^{-1}\sumi}
\newcommand{\sumi}{\sum_{i=1}^{n}}

\newcommand{\osi}{\omega_{*,i}}
\newcommand{\gate}{\textsc{gate}}

\newcommand{\sm}{Supplementary Materials}

\newcommand{\hts}{\hat{\tau}_*}

\newcommand{\necff}{\nef\ cross-fitting}
\newcommand{\necf}{\necff}
\newcommand{\nef}{neighborhood-excluded}

\newcommand{\htf}{Horvitz--Thompson}

\def\begineqs{\begin{equation*}}
\def\endeqs{\end{equation*}}
\def\beginp{\begin{pmatrix}}
\def\endp{\end{pmatrix}}

\def\assm{Assumption}

\newcommand{\lem}{Lemma}

\def\mbr{\mathbb R}

\def\begini{\begin{itemize}}
\def\endi{\end{itemize}}
\def\begine{\begin{enumerate}}
\def\ende{\end{enumerate}}
\def\beginar{\begin{array}}
\def\endar{\end{array}}

\def\E{\mathbb E}

\def\T{\top}

\newcommand{\htau}{\hat\tau}

\newcommand{\ot}[1]{1, \ldots,#1}

\def\beginy{\begin{eqnarray}}
\def\endy{\end{eqnarray}}
\def\begina{\begin{eqnarray*}}
\def\enda{\end{eqnarray*}}
\def\begineqs{\begin{equation*}}
\def\endeqs{\end{equation*}}
\def\begineq{\begin{equation}}
\def\endeq{\end{equation}}

\usepackage{enumerate} 
\usepackage{bm}
\newcommand{\var}{\mathrm{var}}

\newcommand{\den}{\mathrm{dense}}
\newcommand{\dir}{\textsc{dir}}
\newcommand{\ind}{\textsc{ind}}

\newcommand{\ora}{\mathrm{ora}}
\newcommand{\adj}{\mathrm{adj}}
\newcommand{\pr}{\mathbb P}
\newcommand{\argmin}{\mathop{\arg\min}}

\newcommand{\niin}{\mathcal N_{i,\text{in}}}
\newcommand{\niout}{\mathcal N_{i,\text{out}}}
\newcommand{\nis}{\tilde{\mathcal N}_{i}}
\newcommand{\Perp}{\perp\!\!\!\perp}

\newcounter{subassumption}[assumption]
\crefalias{subassumption}{assumption}

\newcommand{\currsubassumptiontag}{}

\newenvironment{subassumptions}
{%
  \begin{list}{}%
  {%
    \setlength{\leftmargin}{2.2em}%
    \setlength{\labelwidth}{2em}%
    \setlength{\labelsep}{0.5em}%
    \setlength{\itemsep}{0.3em}%
    \setlength{\parsep}{0pt}%
    \setlength{\topsep}{0.3em}%
  }%
}
{%
  \end{list}%
}

\newcommand{\assumpitem}[2]{%
  \def\currsubassumptiontag{#1}%
  \refstepcounter{subassumption}%
  \item[(#1)]\label{#2}%
}

\newcommand{\assumpinlinetag}[2]{%
  \def\currsubassumptiontag{#1}%
  \refstepcounter{subassumption}%
  \textup{(#1)}\label{#2}%
}

\newcommand{\bibnospace}{\aftergroup\ignorespaces}

\usepackage[normalem]{ulem}
\begin{document}

\def\spacingset#1{\renewcommand{\baselinestretch}%
{#1}\small\normalsize}

\spacingset{1}


\if1\anon
{
  \title{\bf Unbiased Treatment Effect Estimation under Network Interference via Neighborhood-Excluded Cross-Fitting}
  \author{Haoyang Yu$^{1}$, Anqi Zhao$^{2}$ and Hanzhong Liu$^{1}$\thanks{Address for correspondence: Hanzhong Liu, Department of Statistics and Data Science, Tsinghua University, Beijing, 100084, China. Email: lhz2016@tsinghua.edu.cn. Hanzhong Liu was supported by the Beijing Natural Science Foundation (grant no. F251001), National Natural Science Foundation of China (grant no. 12531012), Shenzhen Science and Technology Program (grant No. AI2026036), and High Performance Computing Center, Tsinghua University.}\vspace{.5cm}\\
    $^{1}$ Department of Statistics and Data Science, Tsinghua University\\
    $^{2}$ Fuqua School of Business, Duke University}
    
    \date{}
  \maketitle
} \fi

\if0\anon
{
  \bigskip
  \bigskip
  \bigskip
  \begin{center}
    {\LARGE\bf Unbiased Treatment Effect Estimation under\\ Network Interference via\vspace{12pt}\\ Neighborhood-Excluded Cross-Fitting}
\end{center}
  \medskip
} \fi

\bigskip
\begin{abstract}
Without interference, cross-fitting enables flexible covariate adjustment while preserving finite-sample unbiasedness under independent unit-level randomization.
Under network interference, out-of-sample prediction alone no longer guarantees unbiasedness: assignments entering evaluation-fold Horvitz–Thompson weights may also affect outcomes in the training sample, inducing dependence between fitted predictions and those weights. 
We develop neighborhood-excluded cross-fitting, which constructs estimand- and design-specific training samples to restore the conditional independence needed for finite-sample unbiasedness without a correctly specified outcome model.
We establish asymptotically valid design-based Wald inference for direct and indirect effects under Bernoulli randomization and for the global average treatment effect under Bernoulli cluster randomization. 
Neighborhood exclusion creates a trade-off in choosing the number of folds: 
unit-level splitting may require the number of folds to grow with average exclusion-neighborhood size, while cluster-level splitting can substantially relaxes this requirement, permitting a fixed number of folds under partial interference. 
For linear adjustment under Bernoulli randomization, 
we derive variance-optimal and confidence-interval-length-optimal procedures,
establish explicit rate conditions allowing the covariate dimension to diverge, and show that the variance-optimal procedure is asymptotically no-harm.
Simulations illustrate the bias from omitting neighborhood exclusion and the precision gains from adjustment. An application to a social network experiment yields confidence intervals substantially shorter than those from the unadjusted estimator.
\end{abstract}

\noindent%
\emph{Keywords:} Covariate adjustment, Design-based inference, High-dimensional covariates, Machine learning, Sample splitting
\vfill

\newpage
\spacingset{1.8} 

\section{Introduction}
\label{sec:intro}

Randomized experiments on networks are increasingly common in economics, the social sciences, and online platforms \citep{sacerdote2001peer,bond201261,paluck2016changing}. 
In these settings, a unit's outcome may depend on both its own treatment and those of its neighbors, violating the stable unit treatment value assumption and complicating causal inference. 
A growing literature develops design-based methods for causal inference under network interference \citep{aronow2017estimating,yu2022estimating,leung2022causal,
gao2025causal,wang2025covariate}.
The design-based perspective treats the potential outcomes, covariates, and network as fixed, 
and grounds inference in the known treatment assignment mechanism rather than a correctly specified outcome model. 

Covariate adjustment is widely used to improve precision in randomized experiments, with well-established benefits in settings without interference \citep{lin2013agnostic,bloniarz2016lasso,lei2021regression,lu2025debiased}.
Recent work extends regression adjustment to network experiments and 
studies its design-based properties under specific interference models  \citep{fan2025causal,gao2025causal,wang2025covariate,lu2024adjusting}.
Two important gaps remain.
(i) \textbf{Finite-sample bias}: 
Standard plug-in adjustments fit the outcome prediction function and estimate the treatment effect using the same data. The fitted predictions can therefore depend on the Horvitz--Thompson weights and induce finite-sample bias.
Under network interference, 
this problem is compounded because assignments entering the Horvitz--Thompson weights may also affect outcomes used to fit the prediction function through the network, creating additional dependence between the predictions and the weights.
(ii) \textbf{High-dimensional covariates}: 
Network experiment can involve rich unit-level, neighborhood, and network-structural covariates, so the covariate dimension may grow with the sample size.
Existing design-based analyses typically assume a fixed covariate dimension and do not provide explicit validity conditions for a diverging dimension.
A general framework combining high-dimensional covariate adjustment, finite-sample unbiasedness, and asymptotically valid design-based inference under network interference is still lacking.

In settings without interference, 
cross-fitting enables covariate adjustment with high-dimensional covariates and flexible prediction methods \citep{chernozhukov2018double,su2023decorrelation,lu2025conditional}. 
The standard procedure partitions the sample into $K$ folds and, for each fold,  predicts outcomes using a model fitted on the other $K-1$ folds, yielding out-of-sample predictions. 
Under independent unit-level assignments with no interference, each unit's fitted prediction is independent of its \htf\ weight conditional on the sample split, preserving finite-sample unbiasedness of the resulting covariate-adjusted estimator \citep{lu2025conditional}. 
Under network interference, however, assignments entering the evaluation-fold weights may also affect outcomes of their neighbors used for training, so fitted predictions can depend on the weights and standard cross-fitting need not be unbiased.

To restore finite-sample unbiasedness, we develop \necf\ (NECF), which constructs training sets so that treatment assignments entering the evaluation-fold \htf\ weights are conditionally independent of those affecting the training outcomes.
The required exclusion rules depend on the estimand and assignment design. 
Under Bernoulli randomization, we exclude first-order out-neighbors for estimating direct effects and second-order neighbors for estimating indirect effects.
Under Bernoulli cluster randomization, we treat clusters as units and apply the unit-level splitting rule with second-order neighborhood exclusion on the induced cluster graph to estimate the global average treatment effect.
These estimand- and design-specific constructions guarantee finite-sample unbiasedness for any prediction algorithm without requiring a correctly specified outcome model.
Under foldwise stability and additional regularity conditions, we further establish oracle equivalence and asymptotically valid design-based Wald inference.

Neighborhood exclusion creates a trade-off in choosing the number of folds $K$.
Smaller $K$ enlarges evaluation sets and may leave little data for training after exclusion.
Larger $K$ retains more observations for training after exclusion but requires stronger control of aggregate prediction error across fits.
For unit-level splitting, our sufficient conditions suggest choosing $K$ on the order of the average degree of the first-order network for direct effects and of the second-order network for indirect effects.
Thus, fixed $K$ suffices when the corresponding average degree is bounded, whereas a diverging average degree may require $K\to\infty$ to retain a nonvanishing training fraction.
Cluster-level splitting can mitigate this trade-off when interference is concentrated within clusters.
Under partial interference, splitting at the level of interference clusters requires no neighborhood exclusion and permits fixed $K$ when the number of clusters grows and no single cluster dominates.
Simulations further show that cluster-level splitting retains substantially more training data than unit-level splitting at a given $K$.

To illustrate the framework, we specialize it to linear adjustment under Bernoulli randomization and derive
variance-optimal and confidence-interval-length-optimal adjustments.
The latter minimizes the expected squared length of Wald confidence intervals exactly for the direct effect and, under suitable conditions, to first order for the indirect effect.
The feasible variance-optimal adjustment is asymptotically no-harm.
For the confidence-interval-length-optimal adjustment, we further show that cluster-level splitting relaxes the requirement on $K$ and 
can permit fixed $K$ under conditions where unit-level splitting requires $K\to\infty$.
We also provide explicit rate conditions allowing the covariate dimension $d$ to diverge with the sample size; in sparse networks with fixed $K$, these conditions reduce to $d=o(n)$.
The framework accommodates existing linear adjustments, such as eigenvector adjustment \citep{lu2026causal}, and extends naturally to flexible machine-learning prediction methods through calibration \citep{cohen2024no}.

\textbf{Related literature.}
Closely related work by \citet{emmenegger2025treatment} develops dependency-graph-based exclusion for cross-fitting with observational network data. 
Our framework differs in three main respects. 
First, we establish design-based inference for randomized experiments with finite-sample unbiasedness, 
whereas \citet{emmenegger2025treatment} study superpopulation inference. 
Second, we tailor the exclusion rule to the estimand and randomization design.
For direct effects, we exclude only first-order out-neighbors of the evaluation units, whereas under neighborhood interference, their dependency-graph-based construction excludes training units within network distance two of the evaluation units. 
Third, our theory explicitly links \(K\) to network connectivity and allows $K$ to diverge with the relevant exclusion-neighborhood size, whereas theirs fixes $K$ and requires a sparse dependency graph. 
We further establish conditions under which cluster-level sample splitting can substantially relax the requirements on $K$ in networks with clustered connectivity.
Relatedly, \citet{yu2026edge} develop a tailored three-fold cross-fitting scheme for edge-level outcomes under dyadic interference. 
We instead target node-level direct, indirect, and global effects under neighborhood interference using estimand-specific exclusion and potentially diverging $K$.

\textbf{Organization.} 
\Cref{sec:framework} presents the framework and neighborhood-excluded cross-fitting. 
\Cref{sec:BRE} develops estimation and inference for direct and indirect effects under Bernoulli randomization. 
\Cref{sec:regression} specializes the theory to linear adjustment and discusses extensions to eigenvector adjustment and flexible machine learning methods via calibration. 
\Cref{sec:CRE} studies inference for the global average treatment effect under Bernoulli cluster randomization. 
Sections \ref{sec:simulation} and \ref{sec:realdata} report simulations and a real-data analysis. 
\Cref{sec:discussion} concludes. 
All proofs are in the \sm.

\textbf{Notation.}
Let $\mathbbm{1}(\cdot)$ denote the indicator function. 
Let $C$ denote a generic positive constant independent of $n$ that may vary across occurrences.
For positive sequences $a_n$ and $b_n$, write $a_n \asymp b_n$ if $cb_n \le a_n \le Cb_n$ for some constants $0 < c \le C < \infty$ and all sufficiently large $n$, and write $a_n\gtrsim b_n$ if $a_n \ge cb_n$ for some constant $c>0$ and all sufficiently large $n$.
For a vector $\bm w=(w_1,\ldots,w_n)^\top$ and an index set $\mathcal S\subseteq\{1,\ldots,n\}$, write $\bm w_{\mathcal S}=(w_j)_{j\in\mathcal S}$ for the subvector indexed by $\mathcal S$. Let $\bm 1_n$ and $\bm 0_n$ denote the $n\times 1$ vectors of ones and zeros. 
Let $z_{1-\alpha/2}$ be the $1-\alpha/2$ quantile of the standard normal distribution, and let $x_{+}=\max(x, 0)$. 

\section{Neighborhood-excluded cross-fitting}
\label{sec:framework}

\subsection{Design-based framework under neighborhood interference}
\label{sec:ni}
Consider a finite population of $n$ units connected by a known directed network with adjacency matrix $\bm A = (A_{ij})\in \{0,1\}^{n\times n}$.
Set $A_{ii}=0$ and, for $i\neq j$, let $A_{ij}=1$ indicate a directed edge from $j$ to $i$, so unit $j$'s treatment may affect unit $i$'s outcome.
For each unit $i$, let $\niin=\{j:A_{ij}=1\}$ denote its \emph{in-neighbors}, whose treatments may affect unit $i$, and let $\niout=\{j:A_{ji}=1\}$ denote its \emph{out-neighbors}, whose outcomes may be affected by $i$'s treatment.
Let $\mm_i=\{i\}\cup\niin$ denote unit $i$'s \emph{assignment neighborhood},  the set of units whose assignments may affect its outcome.
To encode potential dependence from overlapping assignment neighborhoods, define the second-order adjacency matrix $\tilde{\bm A}=(\tilde A_{ij})\in \{0,1\}^{n\times n}$ with $\tilde A_{ii}=0$ and $\tilde A_{ij}=\mathbbm 1(\mm_i\cap\mm_j\neq\varnothing)$ for $i\neq j$. 
Then $\tilde A_{ij}=1$ if and only if $i$ and $j$ are directly connected or share an in-neighbor. 
Let $\nis=\{j:\tilde A_{ij}=1\}$ denote unit $i$'s second-order neighborhood.
By construction, $\tilde{\bm A}$ is symmetric with $\tilde A_{ij}\geq A_{ij}$ and $\niin\cup\niout\subseteq\nis$.
Let $Z_i\in\{0,1\}$ denote the binary treatment assignment for unit $i$, and write $\bm Z=(Z_1,\ldots,Z_n)$.
The distribution of $\bm Z$ defines the assignment mechanism.
\Cref{example.BRE,example.CRE} describe two common designs. 

\begin{example}[Bernoulli randomization]
\label{example.BRE}
Each unit is independently assigned to treatment, with $Z_i\simiid$ Bernoulli($r$) for a common probability $r \in (0,1)$.
\end{example}

\begin{example}[Bernoulli cluster randomization] 
\label{example.CRE}
The $n$ units are partitioned into $M$ clusters.
Each cluster is independently assigned to treatment with common probability
$r\in(0,1)$, and all units within the same cluster receive the corresponding treatment assignment.
\end{example}

For each $\bm z\in\{0,1\}^n$, let $Y_i(\bm z)\in\mathbb R$ denote unit $i$'s potential outcome under $\bm Z = \bm z$. The observed outcome is $Y_i=Y_i(\bm Z)$.
We adopt a design-based perspective: the potential outcomes, covariates, and network are fixed, while randomness arises from the assignment mechanism and sample splitting specified in Sections~\ref{sec:unbiased}--\ref{sec:algorithm}.
\Cref{assumption.neighborhoodinterference} formalizes \emph{neighborhood interference}, under which each unit's outcome depends only on the assignments in its assignment neighborhood.

\begin{assumption}
\label{assumption.neighborhoodinterference}
For any $i \in \{\ot{n}\}$ and $\bm z, \bm z' \in \{0,1\}^n$, $Y_i(\bm z)=Y_i(\bm z')$ if $\bm z_\mmi=\bm z'_{\mmi}$.
\end{assumption}
We consider the direct effect, the indirect effect, and the global average treatment effect (GATE), defined as follows:
\begin{gather*}
    \begin{aligned}
    \tau_{\dir}
    &= \meani \left\{ \E(Y_i\mid Z_i=1) - \E(Y_i\mid Z_i=0) \right\}, \\
    \tau_{\ind}
    &= \meani \sum_{j=1}^n A_{ij}
    \left\{\E(Y_i\mid Z_j=1)-\E(Y_i\mid Z_j=0)\right\},\\
    \tau_{\gate} &= \meani \{Y_i(\bm 1_n) - Y_i(\bm 0_n)\}.
    \end{aligned}
\end{gather*}
All expectations are taken with respect to the assignment mechanism, 
so the direct and indirect effects are design specific.
Under Bernoulli randomization, the direct effect measures the impact of a unit’s own treatment \citep{hudgens2008}, and the indirect effect aggregates marginal spillover effects from each unit's in-neighbors \citep{hu2022average}.
The GATE compares average outcomes under full treatment and full control \citep{ugander2023randomized}.
For $*\in\{\dir,\ind,\gate\}$, a standard estimator for $\tau_*$ is the Horvitz--Thompson estimator $\hat \tau_*=n^{-1}\sum_{i=1}^n \osi Y_i$ \citep{aronow2017estimating, hu2022average}, where
\begin{gather}\label{eq:weights}
\begin{aligned}
\omega_{\dir,i} 
&= \dfrac{Z_i}{\pr(Z_i=1)} - \dfrac{1-Z_i}{\pr(Z_i=0)},\\
\omega_{\ind,i} 
&= \sum_{j=1}^n A_{ij} \left\{ \dfrac{Z_j}{\pr(Z_j=1)} - \dfrac{1-Z_j}{\pr(Z_j=0)} \right\},\\[4pt]
\omega_{\gate,i} 
&= \dfrac{\mathbbm{1}(\bm Z_{\mmi}=\bm 1_{|\mmi|})}{\pr(\bm Z_{\mmi}=\bm 1_{|\mmi|})}
- \dfrac{\mathbbm{1}(\bm Z_{\mmi}=\bm 0_{|\mmi|})}{\pr(\bm Z_{\mmi}=\bm 0_{|\mmi|})}.
\end{aligned}
\end{gather} 
Under \assm~\ref{assumption.neighborhoodinterference} and positivity of the probabilities in \eqref{eq:weights}, 
$\E(\osi Y_i)$ equals the corresponding unit-level contribution to $\tau_*$, so $\hts$ is unbiased.

\subsection{Cross-fitting and finite-sample unbiasedness}
\label{sec:unbiased}
The Horvitz--Thompson estimator is unbiased but may be imprecise. 
For example, under Bernoulli randomization and the conditions of \cite{lu2026causal}, the variance of the indirect-effect estimator can be of order $n\rho_n^2$, where $\rho_n=\sum_{i\neq j}A_{ij}/n^2$ is the network density. 
%
Covariate adjustment can improve precision by removing outcome variation predictable from covariates  \citep{lin2013agnostic,aronow2013class,wager2016high,wu2018loop}.
Given covariates $\bm X_i\in \mathbb R^d$ and a prediction function $f_*:\mathbb{R}^d\to\mathbb{R}$, 
replacing $Y_i$ with the residual $Y_i - f_*(\bm X_i)$ in $\hat{\tau}_*$ yields the adjusted estimator $\meani\osi \{Y_i-f_*(\bm X_i)\} = \hts - \meani \osi f_*(\bm X_i)$.
Because $\E(\osi)=0$, the unadjusted estimator is unbiased for any fixed $f_*$. 
Let $f^\ora_*$ be an oracle prediction function in a prespecified function class, typically defined to minimize a prediction or variance criterion.
In finite-population settings, $f^\ora_*$ generally depends on unobserved potential outcomes and must be estimated from observed data \citep{lin2013agnostic, wager2016high, su2023decorrelation}. 
Estimating $f^\ora_*$ on the same data used for treatment effect evaluation can make the fitted predictions correlated with the \htf\ weights, inducing finite-sample bias.

Cross-fitting separates the data used to fit the prediction function from those used for treatment-effect evaluation \citep{chernozhukov2018double}.
Let $\{\mik\}_{k=1}^K$ denote a random $K$-fold partition of the $n$ units, generated independently of the treatment assignments.
For each fold $k$, choose a training set $\mtsk\subseteq \mikc$, and let
$\mmtsk=\bigcup_{i\in\mtsk}\mmi$
denote the set of units whose assignments may affect the training outcomes in 
$\{Y_i:i\in\mtsk\}$.
Fit $\hfsk$ using outcomes of units in $\mtsk$ and the treatment assignments that may affect these outcomes, namely those of units in $\mmtsk$; covariates of any units may be used.
The resulting cross-fitted estimator is
$$\hat{\tau}_*^{\adj}= n^{-1}\sum_{k=1}^K \sum_{i \in \mik}\omega_{*,i}\hat e_{*,i} = \hts - n^{-1}\sum_{k=1}^K \sum_{i \in \mik} \osi\hfsk(\bm X_i),$$ where $\hat e_{*,i}=Y_i-\hfsk(\bm X_i)$ for $i\in\mik$. 

Standard cross-fitting takes $\mtsk=\mikc$.
Without interference, $\mmi=\{i\}$, so $\oii=0$ and $\ogi=\odi$ depends only on $Z_i$, whereas $\hfsk$ depends only on assignments in $\mmtsk=\mikc$.
Under independent unit-level randomization, $\hfsk$ is therefore conditionally independent of $\{\osi\}_{i\in\mik}$ given the sample split, preserving unbiasedness for arbitrary prediction methods without requiring a correctly specified outcome model \citep{lu2025conditional}.
Under neighborhood interference, however, treatment assignments determining the weights $\{\osi\}_{i\in\mik}$ may also affect outcomes in $\mikc$, so out-of-sample prediction alone no longer guarantees unbiasedness.
To characterize the required separation, let $\mathcal O_{\dir,i}=\{i\}$, $\mathcal O_{\ind,i}=\niin$, and $\mathcal O_{\gate,i}=\mmi$ denote the sets of units whose treatment assignments determine $\osi$, and define $\mosik=\bigcup_{i\in\mik}\mathcal O_{*,i}$.
Conditional on the sample split, the weights $\{\osi\}_{i\in\mik}$ depend only on $\bm Z_{\mosik}$, whereas $\hfsk$ depends only on $\bm Z_{\mmtsk}$.
Consequently, conditional independence of $\bm Z_{\mosik}$ and $\bm Z_{\mmtsk}$ is sufficient for unbiasedness.

\begin{assumption}
\label{assumption.conditionalindependence}
For $*\in\{\dir,\ind,\gate\}$,
(i) $\bm Z_{\mosik}\Perp \bm Z_{\mmtsk}\mid\{\mathcal I_l\}_{l=1}^K$ for every $k\in\{\ot{K}\}$.
(ii) The probabilities in the denominators of $\{\osi\}_{i=1}^n$ in \eqref{eq:weights} are strictly positive.
\end{assumption}

Because the sample split is generated independently of the treatment assignments, the positivity condition in \assm~\ref{assumption.conditionalindependence}(ii) is equivalent to the usual positivity condition under the original randomization distribution.

\begin{theorem}
\label{theorem.conditionalunbiasedness}
Under \Cref{assumption.conditionalindependence,assumption.neighborhoodinterference}, $\E(\hat \tau_{*}^{\adj})=\tau_{*}$ for $* \in \{\dir,\ind,\gate\}$.
\end{theorem}

It thus remains to construct training sets $\mtsk$ that satisfy Assumption~\ref{assumption.conditionalindependence}(i).
Section~\ref{sec:algorithm} achieves this by excluding from each training set units whose outcomes may depend on assignments determining the corresponding evaluation-fold weights.

\begin{remark}
Our residual-based adjustment uses a single prediction function across all treatment states.
Allowing the prediction function to depend on the realized treatment arm or exposure level would generally 
make it depend on $\bm Z$ and 
require an augmented estimator rather than the residual form considered here.
Without interference, \citet{lu2025tyranny} show that appropriately weighted pooled and arm-specific linear adjustments are asymptotically equivalent.
\end{remark}

\begin{figure}[!ht]
    \centering
\newcommand{\hoodfigscale}{0.4}

\pgfmathsetlengthmacro{\NodeSize}{11mm*\hoodfigscale}
\pgfmathsetlengthmacro{\ArrowShorten}{2pt*\hoodfigscale}

\begin{tikzpicture}[
    x=\hoodfigscale cm,
    y=\hoodfigscale cm,
    >=Latex,
    font=\small,
    unit/.style={
        circle,
        draw,
        line width=0.9pt,
        minimum size=\NodeSize,
        inner sep=0pt
    },
    firstout/.style={
        unit,
        fill=yellow!25
    },
    secondonly/.style={
        unit,
        fill=blue!18
    },
    box/.style={
        draw,
        line width=0.8pt
    },
    edge/.style={
        -Latex,
        line width=1pt,
        shorten <=\ArrowShorten,
        shorten >=\ArrowShorten
    }
]

\draw[box] (-1,1) rectangle (2,3);
\node[left] at (-1.5,2) {Evaluation sample $\mik$};

\node[unit] (i) at (0.5,2) {$i$};

\node[firstout]   (j)  at (4.8,3) {$j$};
\node[secondonly] (jp) at (4.8,1) {$j'$};
\node[unit]       (lp) at (9,3) {$\ell'$};
\node[secondonly] (l)  at (9,1) {$\ell$};

\draw[edge] (i)  -- (j);
\draw[edge] (jp) -- (i);
\draw[edge] (jp) -- (l);
\draw[edge] (lp) -- (jp);

\node[above=2pt, align = center] at (j.north)
    {First-order out-neighbor};
\node[below=2pt] at ($(jp.south)!0.5!(l.south)$)
    {Additional second-order neighbors};

\end{tikzpicture}
    \caption{Neighborhood exclusion around $i\in\mik$. Arrows represent directed edges. Yellow marks an out-neighbor of $i$; blue marks two additional second-order neighbors.}
    \label{fig:fail_intro}
\end{figure}
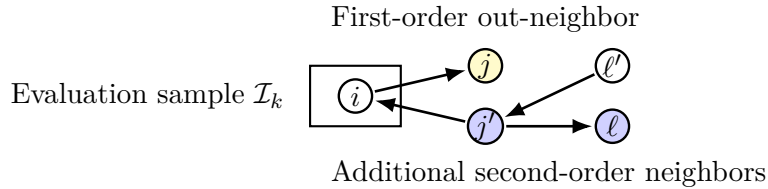

\subsection{Neighborhood-excluded cross-fitting} 
\label{sec:algorithm}
We construct each training set $\mtsk$ to satisfy the non-overlap condition
\begin{equation}\label{eq:nonoverlap}
\mmtsk\cap\mosik=\varnothing,
\qquad
k\in\{\ot K\},
\end{equation}
which ensures Assumption~\ref{assumption.conditionalindependence} under Bernoulli randomization.
Because $\mmtsk\cap\mosik=\cup_{j\in\mtsk, \ i\in\mik}(\mathcal M_j\cap\mathcal O_{*,i})$,
\eqref{eq:nonoverlap} holds if and only if
$\mtsk$ is disjoint from the minimal exclusion set $\mathcal E^{\min}_{*,k}=\{j:\mathcal M_j\cap\mathcal O_{*,i}\neq\varnothing\text{ for some }i\in\mik\}$; any superset is also sufficient.
The minimal exclusion set is the evaluation units together with their first-order out-neighbors for the direct effect, and the evaluation units together with their second-order neighbors for GATE, as illustrated in \Cref{fig:fail_intro}.
For the indirect effect, the minimal exclusion set is the evaluation units and all their second-order neighbors except pure out-neighbors.
%
To avoid a separate exclusion rule for the indirect effect, we use the full second-order neighborhood for both the indirect effect and the GATE. Thus, 
\begin{center}
$\mathcal E_{\dir,k}
=\cup_{i\in\mik}(\{i\}\cup\niout)$, \quad $\mathcal E_{\ind,k}=\mathcal E_{\gate,k}
=\cup_{i\in\mik}(\{i\}\cup\nis)$. 
\end{center}

\begin{algorithm}[!h]
\caption{Neighborhood-excluded cross-fitting for $\htau_*^\adj \ (*\in\{\dir,\ind,\gate\})$}
\label{alg:necf}
\begin{algorithmic}[1]
\REQUIRE
number of folds $K$;
neighborhoods $\{\niout,\nis\}_{i=1}^n$.

\STATE Sample splitting: partition the $n$ units into $K$ folds $\{\mathcal I_k\}_{k=1}^K$ independently of $\bm Z$.

\FOR{$k = 1,\ldots,K$}
    \STATE Define the exclusion sets $\mathcal E_{\dir,k}
=\cup_{i\in\mik}(\{i\}\cup\niout)$, $\mathcal E_{\ind,k}=\mathcal E_{\gate,k}
=\cup_{i\in\mik}(\{i\}\cup\nis)$.
    
    \STATE Construct the training set
    $\mtsk=\{1,\ldots,n\}\setminus\mathcal E_{*,k}$.
    
    \STATE If $\mtsk =\varnothing$, set
$\hfsk=0$. Otherwise, fit $\hfsk$ using the outcomes of units in $\mtsk$, treatment assignments of units in $\mmtsk$, and covariates of any units.
    
    \STATE For each $i\in\mik$, compute the cross-fitted residual $\hat e_{*,i} = Y_i- \hfsk(\bm X_i)$.
\ENDFOR

\STATE \textbf{Output:}
$\hts^{\adj}
= \meani\osi \hat e_{*,i}$.

\end{algorithmic}
\end{algorithm}

\Cref{alg:necf} summarizes the resulting procedure, which enforces the non-overlap condition in \eqref{eq:nonoverlap}.
Under Bernoulli randomization with $r\in(0,1)$, disjoint subvectors of $\bm Z$ are independent and the positivity condition in \assm~\ref{assumption.conditionalindependence}(ii) holds automatically.
Hence, non-overlap is sufficient for \assm~\ref{assumption.conditionalindependence}.
By \thm~\ref{theorem.conditionalunbiasedness}, 
the resulting estimator is therefore unbiased for any sample split generated independently of $\bm Z$.
\Cref{sec:BRE} develops the corresponding inference.
Under designs with dependent treatment assignments, however, disjoint subvectors of $\bm Z$ need not be independent, so non-overlap alone is insufficient.
The sample split and exclusion rule must also account for dependence induced by the randomization design.
\Cref{sec:CRE} develops such a construction under Bernoulli cluster randomization.

\section{Inference under Bernoulli randomization}
\label{sec:BRE}
This section establishes asymptotic normality and conditions for valid Wald inference for the adjusted estimators of the direct and indirect effects under the Bernoulli design in \Cref{example.BRE}.
For the GATE, however, the denominators of the \htf\ weights are the probabilities that all units in the assignment neighborhood $\mmi$ are treated or all are assigned to control, namely $r^{|\mmi|}$ and $(1-r)^{|\mmi|}$.
For fixed $r\in(0,1)$, these probabilities decay exponentially with $|\mmi|$, so the resulting estimator may remain highly variable even after covariate adjustment.
We therefore study inference for the GATE under Bernoulli cluster randomization \citep{ugander2013graph} in \Cref{sec:CRE}.

\subsection{Assumptions on network structure}\label{sec:assm}
Let $m=\sum_{i\neq j} A_{ij}$ denote the number of directed edges and $\rho_n=m/n^2$ the network density. Similarly, let $\tilde\rho_n=n^{-2}\sum_{i\neq j}\tilde A_{ij}$ be the density of the second-order network. Thus, $n\rho_n$ and $n\tilde\rho_n$ are the average degrees of the first- and second-order networks.
Recall that $\niin$ and $\niout$ denote the in- and out-neighborhoods of unit $i$, and let $n_i=\sum_{j=1}^n A_{ij}=|\niin|$.
We impose the following regularity conditions on the network structure.

\begin{assumption}\label{assumption.network}
There exists a constant $C>0$, independent of $n$, such that the following conditions hold as $n\to\infty$:
(i) $\rho_n\to0$;
(ii) $\tilde\rho_n\to0$;
(iii) $\max_i\sum_{j:n_j>0}A_{ji}/n_j\le C$;
(iv) $\max_i|\niin|\le Cn\rho_n$ and $\max_i|\niout|\le Cn\rho_n$; and
(v) $\max_i\sum_{j=1}^n\tilde A_{ij}\le Cn\tilde\rho_n$.
\end{assumption}

\Cref{assumption.network}(i)--(ii) require both the first- and second-order networks to have vanishing densities.
\Cref{assumption.network}(iii) rules out excessive concentration of the normalized exposure contributed by any single unit.
\Cref{assumption.network}(iv)--(v) bound the maximum first-order in- and out-degrees and the maximum second-order degree by constant multiples of their corresponding average degrees, thereby controlling local dependence.
Under \Cref{assumption.network}, we refer to the network as \emph{sparse} if $n\rho_n=O(1)$ and \emph{dense} if $n\rho_n\to\infty$.
The dense regime allows each unit to interact with an increasing number of neighbors and has attracted growing attention in recent work \citep{li2022random,fan2025causal}.
The baseline theory of \citet{lu2026causal}, summarized in \Cref{sec.sm:theory} of the \sm, establishes that $(\hat\tau_*-\tau_*)/\sqrt{\var(\hat\tau_*)}\xrightarrow{d}\mathcal N(0,1)$ for $*\in\{\dir,\ind\}$, with $\var(\hat\tau_{\dir})=O(n^{-1})$ and $\var(\hat\tau_{\ind})=O(n\rho_n^2)$, thereby providing the baseline asymptotic theory for the unadjusted estimators.
We extend this theory to the cross-fitted estimators.

\subsection{Sample splitting strategy and stability condition}
\label{sec:sample.BRE}
Under Bernoulli randomization, unbiasedness follows from Section~\ref{sec:algorithm}. We now specify the sample-splitting rule and examine how neighborhood exclusion and the number of folds $K$ affect training-set retention and prediction stability.

\begin{definition}[Unit-level independent random splitting]
\label{def:unit_splitting}
For a prespecified integer $K \ge 2$,
independently assign each unit to one of $K$ folds with probability $1/K$, yielding a random partition $\mathcal I_1,\ldots,\mathcal I_K$ of $\{1,\ldots,n\}$.
\end{definition}

The number of folds $K$ governs a trade-off between the effective training set size and the stability requirement of the fitted prediction functions.
With neighborhood exclusion, larger evaluation folds generally induce larger exclusion sets, potentially leaving few units for training, especially in dense networks.
Increasing $K$ reduces the size of the evaluation folds and can therefore retain more training observations after exclusion.
This benefit comes at the cost of requiring prediction errors to be controlled uniformly over a larger number of folds.
\Cref{assumption.stability} formalizes this stability requirement.
Let $f_{\dir}^{\ora}$ and $f_{\ind}^{\ora}$ denote the oracle prediction functions for the direct and indirect effects, respectively.

\begin{assumption}
\label{assumption.stability}
As $n\to\infty$, for $*\in\{\dir,\ind\}$, there exist functions $\varepsilon_{*,[k]}:(0,1)\to\mathbb{R}_+$ $(k\in\ot K)$ such that, for every $\delta\in(0,1)$,
(i) $\pr[n^{-1}\sum_{i\in\mik}\{\hfsk(\bm X_i)-f_*^{\ora}(\bm X_i)\}^2\le\varepsilon_{*,[k]}^2(\delta)]\ge1-\delta$ for every $k\in\ot K$; and
(ii) $K\log K\sum_{k=1}^K\varepsilon_{*,[k]}^2\{\delta/(2K)\}\to0$.
\end{assumption}

When $K=O(1)$, \Cref{assumption.stability} reduces to the standard foldwise consistency condition
$n^{-1}\sum_{i\in\mik}\{\hfsk(\bm X_i)-f^{\ora}_*(\bm X_i)\}^2=o_\pr(1)$
for each $k$ \citep{su2023decorrelation,cohen2024no,lu2025conditional}.
This condition is satisfied by a broad class of prediction methods, including linear, nonparametric, and machine learning methods.
When $K\to\infty$, however, \Cref{assumption.stability}(ii) requires the aggregate foldwise prediction error to decay sufficiently fast to offset the $K\log K$ factor, making the stability requirement more stringent.
Section~\ref{sec:regression} provides explicit sufficient conditions under linear adjustment.

The required number of folds depends on the size of the exclusion neighborhoods.
Under \Cref{assumption.network}(iv), each unit's first-order degree is bounded by $Cn\rho_n$.
Consequently, under independent unit-level sample splitting, $K$ must be sufficiently large to retain a nonvanishing training set after neighborhood exclusion.
For example, taking
$K=\max\{2,\lceil(C+1)(1+n\rho_n)\rceil\}$
ensures that the expected fraction of units excluded from each training set is at most $C/(C+1)$.
Lemma~\ref{lem:training-size} strengthens this result by showing that
$\min_{1\le k\le K}|\mathcal T_{*,k}|/n$
is bounded away from zero with probability tending to one.
Accordingly, when $n\rho_n=O(1)$, one may take $K=O(1)$, so that only the fixed-$K$ version of \Cref{assumption.stability} is required.
By contrast, when $n\rho_n\to\infty$, retaining a nonvanishing training fraction may require $K\to\infty$, thereby strengthening the stability requirement in \Cref{assumption.stability}.
The same argument applies to second-order neighborhood exclusion under \Cref{assumption.network}(v), with $n\rho_n$ replaced by $n\tilde\rho_n$.
This trade-off can be mitigated by cluster-level splitting based on a prespecified partition of the units, as described in \Cref{def:cluster_splitting_BRE}.

\begin{definition}[Cluster-level independent random splitting]
\label{def:cluster_splitting_BRE}
Given a partition of the units that is fixed before sample splitting, and an integer $K\ge2$, assign each cluster independently and uniformly at random to one of the $K$ folds. All units within the same cluster are assigned to the same fold.
\end{definition}

The advantage of cluster-level splitting is most pronounced under partial interference \citep{hudgens2008}, where interference is confined within clusters.
If the partition coincides with the interference clusters, the training and evaluation sets do not interfere, so no neighborhood exclusion is required.
If the number of clusters grows and no single cluster dominates the sample, any fixed $K$ retains a nonvanishing training fraction with probability tending to one.
More generally, when most links lie within clusters, as in stochastic block models, cluster-level splitting induces fewer connections between the training and evaluation sets than unit-level splitting.
Consequently, neighborhood exclusion removes fewer units, allowing a smaller $K$ while retaining a nonvanishing training fraction; see \Cref{sec.sm:clusterversusunit} for a numerical illustration.
The choice of $K$ therefore trades off the training set retained after neighborhood exclusion against the stability requirement for prediction.
Cluster-level splitting, when available, relaxes this trade-off.

\subsection{Inference for covariate-adjusted estimators}
\label{sec:inference}

We now establish asymptotic normality and variance estimation for the covariate-adjusted estimators under Bernoulli randomization.
Under \Cref{assumption.neighborhoodinterference}, each potential outcome $Y_i(\bm z)$ admits the exact decomposition: $Y_i(\bm z) = \sum_{\mathcal S\subseteq \niin}\{ \alpha_{i,\mathcal S}\prod_{j\in\mathcal S}(z_j-r) + \alpha_{i,\{i\}\cup\mathcal S}(z_i-r)\prod_{j\in\mathcal S}(z_j-r)\}$, where $\alpha_{i,\mathcal U} = \sum_{\bm z_{\mathcal U}\in\{0,1\}^{|\mathcal U|}} \{ \prod_{j\in\mathcal U}(2z_j-1) \} \E(Y_i \mid \bm Z_{\mathcal U}=\bm z_{\mathcal U})$ for any $\mathcal U \subseteq \{i\}\cup\niin$  \citep{lu2026causal}. 
For $* \in \{\dir,\ind\}$, let $e_{*,i}=Y_i-f_*^{\ora}(\bm X_i)$ be the oracle residual, and let $\alpha_{*,i,\mathcal U}$ denote the analog of $\alpha_{i,\mathcal U}$ with $Y_i$ replaced by $e_{*,i}$. Let $\hat{\tau}_*^{\ora}= n^{-1}\sum_{k=1}^K \sum_{i \in \mik}\osi  e_{*,i}$.

\begin{assumption}
\label{assumption.bounded}
There exists a constant $C>0$, independent of $n$, such that for every $\bm z\in\{0,1\}^n$:
(i) $\max_{1\le i\le n}|Y_i(\bm z)|\le C$; and
(ii) $n_i|Y_i(\bm z)-Y_i(\bm z^{(j)})|\le C$ for every $i$ with $n_i>0$ and every $j\neq i$, where $\bm z^{(j)}\in\{0,1\}^n$ differs from $\bm z$ only in the treatment of unit $j$.
\end{assumption}

\Cref{assumption.bounded} uniformly bounds the potential outcomes and the influence of any single neighbor's treatment on the outcome, thereby ruling out single influential neighbors.

\begin{assumption}
\label{assumption.oracle}
There exists a constant $C>0$, independent of $n$, such that for each $*\in\{\dir,\ind\}$ and all $i=1,\ldots,n$, $
    \max \{
    |\alpha_{*,i,\varnothing}|,
    |\alpha_{*,i,\{i\}}|,
    \max_{\substack{1\le k\le n_i, |\mathcal S|=k, \mathcal S\subseteq \niin}}
        (n_i^k/k!)|\alpha_{*,i,\mathcal S}|, $ $
    \max_{\substack{1\le k\le n_i, |\mathcal S|=k, \mathcal S\subseteq \niin}}
        (n_i^k/k!)|\alpha_{*,i,\{i\}\cup\mathcal S}|
    \}
    \le C .$
In addition, assume that:

\begin{subassumptions}

\assumpitem{\dir}{assumption.oracle.direct}
$\liminf_{n} n\var(\hat \tau_{\dir}^{\ora})>0$;\quad
\assumpinlinetag{\ind}{assumption.oracle.indirect}
(i) $\liminf_{n} \var(\hti^\ora)/(n\rho_n^2)>0$, (ii) there exists $C_+>0$ such that $n\rho_n\geq C_+$, and (iii) $n\rho_n=O(1)$ or $n\rho_n\to \infty$.

\end{subassumptions}

\end{assumption}

The common bound on $\alpha_{*,\cdot}$ in \Cref{assumption.oracle} ensures that \Cref{assumption.bounded} continues to hold after replacing $Y_i(\bm z)$ by $Y_i(\bm z)-f_*^{\ora}(\bm X_i)$.
The direct- and indirect-effect conditions then impose nondegeneracy at the appropriate variance scale for the corresponding estimands.
The remaining conditions for the indirect effect rule out vanishing average degree and require $n\rho_n$ to lie in either the bounded-degree or the diverging-degree regime.

\begin{theorem}
\label{theorem.clt}
Suppose treatment is assigned according to the Bernoulli randomization.
(i) Under Assumptions \ref{assumption.neighborhoodinterference}, \ref{assumption.conditionalindependence}, \ref{assumption.network}(i), \ref{assumption.network}(iii), \ref{assumption.stability}, and \ref{assumption.oracle.direct},
$(\hat{\tau}_{\dir}^{\adj}-\tau_{\dir})/\sqrt{\var(\hat{\tau}_{\dir}^{\ora})}\xrightarrow{d}\mathcal N(0,1)$.
(ii) Under Assumptions \ref{assumption.neighborhoodinterference}--\ref{assumption.stability} and \ref{assumption.oracle.indirect},
$(\hat{\tau}_{\ind}^{\adj}-\tau_{\ind})/\sqrt{\var(\hat{\tau}_{\ind}^{\ora})}\xrightarrow{d}\mathcal N(0,1)$. 
\end{theorem}

To use \Cref{theorem.clt} for inference, we need variance estimators.
We start from existing design-based variance estimators for the unadjusted Horvitz--Thompson estimator.
For sparse networks, a natural choice is the dependency-graph estimator of \citet{aronow2017estimating}, denoted by $\hat V_{\mathrm{AS},*}$:
$\hat V_{\mathrm{AS},*}=n^{-2}\sum_{i=1}^n\sum_{j\in\tilde{\mm}_i}B_{ij}\osi \omega_{*,j}Y_iY_j$, where $\tilde{\mm}_i=\{i\}\cup\nis$, and $B_{ij}$ is determined by the distribution of $\bm Z$. Explicit expressions for $B_{ij}$ are provided in \Cref{sec.sm:AS} in the \sm.
In dense networks, however, this estimator can be overly conservative, attaining orders as large as $n\rho_n^2$ for $\hat\tau_{\dir}$ and $n^2\rho_n^3$ for $\hat\tau_{\mathrm{ind}}$. Relative to the true variance rates $n^{-1}$ and $n\rho_n^2$, respectively, these correspond to inflation multiples of $(n\rho_n)^2$ and $n\rho_n$, producing unnecessarily wide confidence intervals.
To address this issue, \citet{lu2026causal} propose the following variance estimators for dense-network settings:
\beginy\label{eq:hv_dense}
    \hat V_{\den,\dir}=\dfrac 1{n^2}\sum_{i=1}^nY_i^2\left(\dfrac{Z_i}{r_1^2}+\dfrac{1-Z_i}{r_0^2}\right), \ 
    \hat V_{\den,\ind}=\dfrac 1{n^2}\sum_{i=1}^n\left(\sum_{j=1}^nA_{ji}Y_j\right)^2\left(\dfrac{Z_i}{r_1^2}+\dfrac{1-Z_i}{r_0^2}\right),
\endy
where $r_1=r$ and $r_0=1-r_1$.
For the proposed covariate-adjusted estimators, we use similar variance estimators, with outcomes replaced by residuals.
Specifically, let $\hat V^{\ora}_{\dagger,*}$ and $\hat V^{\adj}_{\dagger,*}$ be obtained by replacing $Y_i$ with $e_{*,i}$ and $\hat e_{*,i}$, respectively, where $\dagger \in \{\mathrm{AS}, \den\}$ and $* \in \{\dir,\ind\}$.
\Cref{theorem.variance} summarizes the corresponding properties for $\hat V^{\adj}_{\mathrm{AS},*}$ in sparse networks and $\hat V^{\adj}_{\den,*}$ in dense networks.

\begin{theorem}
\label{theorem.variance}
Under Assumptions \ref{assumption.neighborhoodinterference}--\ref{assumption.bounded} and \ref{assumption.oracle}($*$) for $*\in\{\dir,\ind\}$,
(i) if $n\rho_n=O(1)$, $\hat V^{\adj}_{\mathrm{AS},*}-\var(\hat\tau_*^{\ora})=\Delta_{\mathrm{AS},*}+o_\pr\{\var(\hat\tau_*^{\ora})\}$, where $\Delta_{\mathrm{AS},*}=\E(\hat V^{\ora}_{\mathrm{AS},*})-\var(\hat\tau_*^{\ora})\ge0$;
(ii) if $n\rho_n\to\infty$,  $\hat V^{\adj}_{\den,\dir}-\var(\hat\tau_{\dir}^{\ora})=\Delta_{\den,\dir}+o_\pr\{\var(\hat\tau_{\dir}^{\ora})\}$, where $\Delta_{\den,\dir}=\E(\hat V^{\ora}_{\den,\dir})-\var(\hat\tau_{\dir}^{\ora})$; and
(iii) if $n\rho_n\to\infty$, $\hat V^{\adj}_{\den,\ind}-\var(\hti^\ora)=o_\pr\{\var(\hti^\ora)\}$.
Explicit formulas for $\Delta_{\mathrm{AS},*}$ and $\Delta_{\den,\dir}$ are given in \lem~\ref{lem:deltas} of the \sm.
\end{theorem}

\Cref{theorem.variance} establishes asymptotically conservative inference for $\htau_*^{\adj}$ ($*\in\{\dir,\ind\}$) in sparse networks using $\hat V^{\adj}_{\mathrm{AS},*}$, and asymptotically exact inference for $\htau_{\ind}^{\adj}$ in dense networks using $\hat V^{\adj}_{\den,\ind}$.
For the direct effect in dense networks, however, $\Delta_{\den,\dir}$ need not be nonnegative; valid inference therefore requires the additional condition in \Cref{cor:coverage}(ii), for which \citet{lu2026causal} provide sufficient conditions.

\begin{corollary}[Coverage of Wald Intervals]\label{cor:coverage}
For $\alpha\in(0,1)$, $\dagger\in\{\mathrm{AS},\den\}$, and $*\in\{\dir,\ind\}$, let
$\mathrm{CI}_{\dagger,*}=[\hat\tau_*^{\adj}\pm z_{1-\alpha/2}(\hat V_{\dagger,*}^{\adj})_+^{1/2}]$.
Under Assumptions \ref{assumption.neighborhoodinterference}--\ref{assumption.bounded} and \ref{assumption.oracle}($*$) for $*\in\{\dir,\ind\}$,
(i) if $n\rho_n=O(1)$, then
$\liminf_{n\to\infty}\pr(\tau_*\in\mathrm{CI}_{\mathrm{AS},*})\ge1-\alpha$
for $*\in\{\dir,\ind\}$;
(ii) if $n\rho_n\to\infty$ and
$\liminf_{n\to\infty}\Delta_{\den,\dir}/\var(\hat\tau_{\dir}^{\ora})\ge0$,
then
$\liminf_{n\to\infty}\pr(\tau_{\dir}\in\mathrm{CI}_{\den,\dir})\ge1-\alpha$;
and
(iii) if $n\rho_n\to\infty$, then
$\pr(\tau_{\ind}\in\mathrm{CI}_{\den,\ind})\to1-\alpha$.
\end{corollary}

\section{Application to linear adjustment}
\label{sec:regression}

The results in \Cref{sec:BRE} rely on \Cref{assumption.stability}, which is formulated for general prediction methods.
This section specializes it to linear adjustment in Bernoulli randomization.
Let $\bm X=(\bm X_1,\ldots,\bm X_n)^\top\in\mbr^{n\times d}$ denote the covariate matrix.
For $\bb\in\mbr^d$, define the linearly adjusted estimator
$\hts^\lr(\bb)=n^{-1}\sum_{i=1}^n\osi(Y_i-\bm X_i^\top\bb)$.
We study two oracle coefficients: $\bsorav$, which minimizes the variance of $\hts^\lr(\bb)$, and $\bsoraci$, which minimizes the expected squared length of the dense-network Wald confidence interval, and derive explicit sufficient conditions for \assm~\ref{assumption.stability} in terms of $(d, K, \rho_n, \tilde \rho_n)$.

\subsection{Variance-optimal linear adjustment}

We first define the oracle coefficient $\bsorav = \argmin_{\bm \beta} \var\{\hat{\tau}_*^{\mathrm{LR}}(\bm \beta)\}$ to minimize the variance of the linearly adjusted estimator.
Solving this quadratic minimization problem yields the unique oracle coefficients given in \Cref{theorem.donoharm}.

\begin{theorem}
\label{theorem.donoharm}
Under Assumptions \ref{assumption.neighborhoodinterference}, if $\bm X^{\top}\bm X$ is nonsingular, then 
\begineqs
\bdorav=
\left(\dfrac{\bm X^{\top}\bm X}{r_1 r_0}\right)^{-1}
\sum_{i=1}^n\left\{\bm X_i \E(\omega_{\dir,i}^2 Y_i)+\sum_{j=1}^n A_{ij}\bm X_j \E(\omega_{\dir,i}\omega_{\dir,j} Y_i)\right\}.
\endeqs
Under Assumptions \ref{assumption.neighborhoodinterference}, if $\bm X^{\top}\bm A\bm A^{\top}\bm X$ is nonsingular, then 
\begineqs
\biorav=
\left(\dfrac{\bm X^{\top}\bm A\bm A^{\top}\bm X}{r_1 r_0}\right)^{-1}
\sum_{i=1}^n\left\{\bm X_i \E(\omega_{\ind,i}^2 Y_i)+\sum_{j=1}^n \tilde A_{ij}\bm X_j \E(\omega_{\ind,i}\omega_{\ind,j} Y_i)\right\}.
\endeqs
For $*\in\{\dir,\ind\}$, $\var\{
\hts^{\mathrm{LR}}(\bsorav)\}
\le
\var(\hts)$.
\end{theorem}

\thm~\ref{theorem.donoharm} establishes finite-sample no-harm for the oracle adjustment.
Let $R_{*,i,[k]}\in\{0,1\}$ indicate whether unit $i$ is included in the training set $\mtsk$ for fold $k$.
For $*\in\{\dir,\ind\}$, define the fold-specific inverse-probability-weighted estimator of $\bsorav$ by
\begin{align}
\label{eq:direct1}
&&\hbdkv
=
\left(\dfrac{\bm X^{\top}\bm X}{r_1 r_0}\right)^{-1}
\sum_{i=1}^n\left(\gamma_{\dir,i,[k]}\omega_{\dir,i}^2\bm X_i+\sum_{j=1}^n \gamma_{\dir,ij,[k]}\omega_{\dir,i}\omega_{\dir,j}A_{ij} \bm X_j\right) Y_i,
\end{align}
\begin{align}
\label{eq:indirect1}
&&\hbikv = \left(\dfrac{\bm X^{\top}\bm A\bm A^{\top}\bm X}{r_1 r_0}\right)^{-1}\sum_{i=1}^n\left(\gamma_{\ind,i,[k]}\omega_{\ind,i}^2 \bm X_i+\sum_{j=1}^n \gamma_{\ind,ij,[k]}\tilde A_{ij}\omega_{\ind,i}\omega_{\ind,j} \bm X_j\right) Y_i,
\end{align}
where $\gamma_{*,i,[k]} = R_{*,i,[k]}/\E(R_{*,i,[k]})$ and $\gamma_{*,ij,[k]} = R_{*,i,[k]}R_{*,j,[k]}/\E(R_{*,i,[k]}R_{*,j,[k]})$ correct for neighborhood exclusion, so $\hbsk^\textup{V}$ is unbiased for $\bsorav$.
The resulting variance-optimal treatment effect estimator is $\hat \tau_*^{\mathrm{LR}\text{-}\mathrm{V}}=n^{-1}\sum_{k=1}^K\sum_{i\in\mik}\osi (Y_i-\bm X_i^{\top}\hbsk^\textup{V})$.
We next state the sufficient conditions for \Cref{assumption.stability}.

\begin{assumption}
\label{assumption.R}
There exists a constant $\epsilon\in(0,1)$ such that
$\pr(i\in\mtsk)\ge\epsilon$
for every fold $k$ and every unit $i$.
\end{assumption}

\Cref{assumption.R} requires every unit to remain in the training set with probability bounded away from zero for each fold.
Under unit-level splitting in \Cref{def:unit_splitting}, a unit is retained for direct-effect estimation only if neither the unit nor any of its in-neighbors is assigned to the evaluation fold, giving probability $(1-K^{-1})^{n_i+1}$.
For indirect-effect estimation, the corresponding probability is of order $(1-K^{-1})^{\tilde n_i+1}$ under second-order neighborhood exclusion, where $\tilde n_i=\sum_{j=1}^n\tilde A_{ij}$ denotes the second-order degree of unit $i$.
Consequently, Assumption~\ref{assumption.R} holds under unit-level splitting whenever $K\gtrsim1+n\rho_n$ for the direct effect or $K\gtrsim1+n\tilde\rho_n$ for the indirect effect.
Under cluster-level splitting in \Cref{def:cluster_splitting_BRE}, the corresponding retention probability is of order $(1-K^{-1})^{q_i}$, where $q_i$ is the number of clusters whose assignment to the evaluation fold would exclude unit $i$ from the training set.
Accordingly, if $q_i$ is uniformly bounded or grows sufficiently slowly, cluster-level splitting permits slower growth of $K$, consistent with the discussion in \Cref{sec:sample.BRE}.

For a symmetric matrix $\bm H$, let $\lambda_{\max}(\bm H)$ and $\lambda_{\min}(\bm H)$ denote its largest and smallest eigenvalues, respectively.
\Cref{assumption.X} below bounds the influence of each unit on both the ordinary and network-transformed least-squares fits.

\begin{assumption}
\label{assumption.X}
The entries of $\bm X$ are uniformly bounded, $\lambda_{\max}(\bm X^\top\bm X)=O(n)$, and:
\begin{subassumptions}

\assumpitem{\dir}{assumption.X.direct}
$\bm X^\top \bm X$ is nonsingular and $\maxi \|(\bm X^\top\bm X)^{-1}\bm X_i\|_2=O(\sqrt d/n)$.

\assumpitem{\ind}{assumption.X.indirect}
$\bm X^\top \bm A \bm A^\top \bm X$ is nonsingular and $\maxi\|(\bm X^\top\bm A\bm A^\top\bm X)^{-1}\bm X_i\|_2=O\{\sqrt d/(n^3\rho_n^2)\}$.
\end{subassumptions}
\end{assumption}

Uniform boundedness implies $\max_i\|\bm X_i\|_2=O(\sqrt d)$.
Consequently, the direct-effect condition yields
$h_{ii}=\bm X_i^\top(\bm X^\top\bm X)^{-1}\bm X_i
\le
\|\bm X_i\|_2\|(\bm X^\top\bm X)^{-1}\bm X_i\|_2
=
O(d/n)$,
so each unit's leverage is of the same order as the average leverage and no single unit dominates the least-squares fit.
A sufficient condition is $\lambda_{\min}(\bm X^\top\bm X)\gtrsim n$.
The indirect-effect condition has the same interpretation for the network-transformed covariates $\bm A^\top\bm X$, and is implied by
$\lambda_{\min}(\bm X^\top\bm A\bm A^\top\bm X)\gtrsim n^3\rho_n^2$.

\begin{theorem}
\label{theorem.regression}
Consider the linear functions $\hfsk(\bm X_i)=\bm X_i^{\top}\hbskv$ under the unit-level independent splitting in Definition~\ref{def:unit_splitting}.
For the direct effect, under Assumptions \ref{assumption.neighborhoodinterference}--\ref{assumption.network}, \ref{assumption.bounded}, \ref{assumption.oracle.direct}, \ref{assumption.R}, and \ref{assumption.X.direct}, \Cref{assumption.stability} holds if $d (1+n\rho_n)^2 K^2(\log K)^2/n\to0$.
For the indirect effect, under Assumptions \ref{assumption.neighborhoodinterference}--\ref{assumption.network}, \ref{assumption.bounded}, \ref{assumption.oracle.indirect}, \ref{assumption.R}, and \ref{assumption.X.indirect}, \Cref{assumption.stability} holds if $d (1+n\tilde\rho_n)^2 K^2(\log K)^2/n\to0$.
\end{theorem}

Together with Theorem~\ref{theorem.crossfitting}, Theorem~\ref{theorem.regression} implies that the feasible cross-fitted estimator is first-order equivalent to its oracle counterpart and therefore asymptotically inherits its no-harm guarantee.
In sparse networks with $n\rho_n=O(1)$, the dimensionality requirement simplifies to $d=o(n)$, comparable to conditions in recent work on high-dimensional covariate adjustment without interference \citep{su2023decorrelation,lu2025debiased}.

In settings without interference, ordinary least-squares adjustment can simultaneously reduce the variance of the point estimator and the expected squared length of Wald confidence intervals \citep{lin2013agnostic,li2017general}.
Under network interference, however, the variance-optimal adjustment need not minimize the expected squared confidence-interval length.
The next subsection therefore considers this objective separately.

\subsection{Dense-network CI-length-optimal linear adjustment}

Let $\hat V^{\mathrm{LR}}_{\den,*}(\bm\beta)$ be the dense-network variance estimator of $\hts^\lr(\bm \beta)$, with $Y_i$ in \eqref{eq:hv_dense} replaced by $\tyi = Y_i - \bm X_i^\T\bm \beta$. 
We next choose the adjustment to minimize the expected squared length of Wald intervals based on $\hat V^{\mathrm{LR}}_{\den,*}(\bm\beta)$. 
%
For the direct effect, define $\bdoraci= \argmin_{\bm \beta} \E\{\hat V^{\mathrm{LR}}_{\den,\dir}(\bm\beta)\}$, with a unique minimizer $\bdoraci = (\bm X^{\top}\bm X)^{-1}\sum_{i=1}^n \E[\{(r_0/r_1)Z_i+(r_1/r_0)(1-Z_i)\}\bm X_iY_i]$. 
The weight $(r_0/r_1)Z_i+(r_1/r_0)(1-Z_i)$ coincides with the weight in tyranny-of-the-minority regression under no interference \citep{lin2013agnostic,lu2025tyranny}. 
The fold-specific inverse-probability-weighted estimator of $\bdoraci$ is given by
\begineq
\label{eq:direct2}
\hbdkci
= \left(\bm X^{\top}\bm X\right)^{-1}\sum_{i=1}^n\dfrac{R_{\dir,i,[k]}}{\E(R_{\dir,i,[k]})}\left\{\dfrac{r_0}{r_1}Z_i+\dfrac{r_1}{r_0}(1-Z_i)\right\}\bm X_iY_i.
\endeq
For the indirect effect, applying \citet{lu2026causal} to $\tyi$ yields
$\E\{\hat V^\lr_{\den,\ind}(\bb)\}=Q_n(\bb)+O(\rho_n)$
for every fixed $\bb$ satisfying $\max_i|\tyi|=O(1)$, where
$Q_n(\bb)=n^{-2}(r_1r_0)^{-1}\E\{\tby\}^\top\bm A\bm A^\top\E\{\tby\}$ with
$\tby=(\tilde Y_1(\bb),\ldots,\tilde Y_n(\bb))^\top$.
If $\bm X^\top\bm A\bm A^\top\bm X$ is nonsingular, the unique minimizer of $Q_n(\bb)$ is
$\bioraci=(\bm X^\top\bm A\bm A^\top\bm X)^{-1}\sum_{i=1}^n\sum_{j=1}^n n_{ij}\bm X_j\E(Y_i)$,
where $n_{ij}=\sum_{k=1}^nA_{ik}A_{jk}$.
Furthermore, if $n\rho_n\to\infty$ and the covariates do not fully explain the network-aggregated means in the sense that
$\|\bm A^\top\E\{\tilde{\bm Y}(\bioraci)\}\|_2^2\gtrsim n(n\rho_n)^2$,
then $Q_n(\bb)\ge Q_n(\bioraci)\gtrsim n\rho_n^2$, so the $O(\rho_n)$ remainder is asymptotically negligible and $\bioraci$ minimizes the expected squared confidence-interval length to first order.
Moreover, $\var\{\hti^\lr(\bb)\}=Q_n(\bb)+O(\rho_n)$, implying that $\bioraci$ also minimizes the variance to first order.
It need not coincide with the exact variance-optimal coefficient $\biorav$ in \thm~\ref{theorem.donoharm}, because lower-order terms may alter the minimizer.
The fold-specific inverse-probability-weighted estimator of $\bioraci$ is
\begineq
\label{eq:indirect2}
\hbikci
= \left(\bm X^{\top}\bm A\bm A^{\top}\bm X\right)^{-1}
\sum_{i=1}^n \dfrac{R_{\ind,i,[k]}}{\E(R_{\ind,i,[k]})}
\sum_{j=1}^n n_{ij}\bm X_j Y_i.
\endeq
From \eqref{eq:direct2}--\eqref{eq:indirect2}, define the dense-network CI-length-optimal treatment effect estimators as $\hat \tau_*^{\mathrm{LR}\text{-}\mathrm{CI}}=n^{-1}\sum_{k=1}^K\sum_{i\in\mik}\osi (Y_i-\bm X_i^{\top}\hbskci)$.

\begin{theorem}
\label{theorem.regression.CI}
Consider the linear functions $\hfsk(\bm X_i)=\bm X_i^{\top}\hbskci$ 
under the unit-level independent splitting in Definition~\ref{def:unit_splitting}.
For $*\in\{\dir,\ind\}$, under Assumptions \ref{assumption.neighborhoodinterference}--\ref{assumption.network}, \ref{assumption.bounded}, \ref{assumption.oracle}($*$), \ref{assumption.R}, and \ref{assumption.X}($*$), \Cref{assumption.stability} holds if $dK^2(\log K)^2/n\to0$.
\end{theorem}

Compared with the variance-optimal adjustment in Theorem~\ref{theorem.regression}, Theorem~\ref{theorem.regression.CI} removes the explicit exclusion-degree factor from the rate condition.
This improvement reflects that the CI-length-optimal coefficients in \eqref{eq:direct2}--\eqref{eq:indirect2} depend only on single-unit inclusion-weighted aggregates, whereas the variance-optimal coefficients in \eqref{eq:direct1}--\eqref{eq:indirect1} additionally involve pairwise inclusion weights and network-induced cross-unit terms.

Although the rate condition in Theorem~\ref{theorem.regression.CI} has no explicit exclusion-degree factor, the network still enters through the number of folds under unit-level splitting.
Let $D_{\dir,n}=1+n\rho_n$ and $D_{\ind,n}=1+n\tilde\rho_n$ denote the average sizes of the direct- and indirect-effect exclusion neighborhoods, respectively.
Under Assumption~\ref{assumption.network}, choosing $K\asymp D_{*,n}$ ensures Assumption~\ref{assumption.R}; substituting this choice into Theorem~\ref{theorem.regression.CI} gives $dD_{*,n}^2(\log D_{*,n})^2/n\to0$ when $D_{*,n}\to\infty$, whereas the condition reduces to $d=o(n)$ when $D_{*,n}=O(1)$.

Cluster-level splitting can weaken this requirement by replacing unit-level exclusion degrees with cluster-level overlap measures.
Under partial interference, with splitting clusters equal to interference clusters, a sufficient condition is $dc_{\max}/n\to0$, where $c_{\max}$ is the largest cluster size.
This condition reduces to $d=o(n)$ for uniformly bounded cluster sizes and to $d=o(M)$ for $M$ balanced clusters.
Beyond partial interference, the corresponding condition depends on the cluster sizes, the number of splitting clusters intersecting each exclusion neighborhood, and the aggregate overlap among these cluster-level exclusion sets; see Theorem~\ref{thm:cluster-LRL} in the \sm.

\begin{remark}
The linear adjustment framework can also be applied after augmenting or transforming the covariates.
For example, eigenvector (EV) adjustment \citep{lu2026causal} adds leading eigenvectors of $\bm A\bm A^{\top}$ to capture network-level variation when this matrix has a few dominant eigenvalues.
In addition, fitted values from a flexible machine learning model can serve as covariates in the final linear adjustment.
The resulting calibrated estimator remains a linear adjustment based on an augmented covariate vector. Although this augmented covariate vector is random, the variance-reduction guarantee may be established under additional conditions using the proof technique of \citet{cohen2024no}.
\end{remark}

\section{Inference for GATE under cluster randomization}
\label{sec:CRE}

Building on the preceding analysis of direct and indirect effects under Bernoulli randomization, we now consider the GATE under Bernoulli cluster randomization, which avoids the exponentially small exposure probabilities that arise under unit-level randomization.  We state the main inferential result here and defer the full construction, assumptions, and intermediate results to \Cref{sec.sm:CRE.full} of the \sm.

Let $\mathcal C_1,\ldots,\mathcal C_M$ be the clusters, with assignments $Z_{[m]}\simiid\mathrm{Bernoulli}(r)$.  For unit $i$, let $\mg_{i,\mathrm{in}}$ contain its own cluster and every cluster containing an in-neighbor.  Its weight is $\omega_{\gate,i}=\prod_{\ell\in\mg_{i,\mathrm{in}}}(Z_{[\ell]}/r)-\prod_{\ell\in\mg_{i,\mathrm{in}}}\{(1-Z_{[\ell]})/(1-r)\}$, and the adjusted estimator is $\hat\tau_{\gate}^{\adj}=n^{-1}\sum_{i=1}^n\omega_{\gate,i}\hat e_{\gate,i}$.
For sample splitting, treat the $M$ clusters as units and apply the unit-level splitting scheme in \Cref{def:unit_splitting}: assign clusters independently to folds and exclude from training all clusters in the second-order neighborhoods of the evaluation clusters. When these neighborhoods are uniformly bounded, the number of folds may remain fixed.

Treating clusters as units simplifies sample splitting but does not make GATE inference immediate.  The GATE contrasts full treatment with full control rather than a marginal treatment contrast under the Bernoulli design, and hence has no direct counterpart among the estimands analyzed in \Cref{sec:BRE}.  Moreover, the target remains unit-weighted, clusters may differ in size, and units within the same cluster may have different exposure neighborhoods.  Establishing asymptotic normality and valid variance estimation therefore requires a separate dependency-graph argument.  


Let $\tilde{\bm A}$ denote the second-order adjacency matrix of the cluster-expanded graph (see \Cref{sec.sm:CRE.full} of the \sm\ for details), and let $\hat e_{\gate,i}=Y_i-\hat f_{\gate,[k]}(\bm X_i)$ for $i\in\mik$.  The feasible variance estimator is $\hat V_{\gate}^{\adj}=n^{-2}\sum_{i=1}^n\sum_{j=1}^n\{\tilde A_{ij}+\mathbbm 1(i=j)\}(\omega_{\gate,i}\hat e_{\gate,i}-\hat\tau_{\gate}^{\adj})(\omega_{\gate,j}\hat e_{\gate,j}-\hat\tau_{\gate}^{\adj})$.
Let $(\hat V_{\gate}^{\adj})_+=\max\{\hat V_{\gate}^{\adj},0\}$. Let $e_{\gate,i}=Y_i-f_{\gate}^{\ora}(\bm X_i)$ be the oracle residual, and let  $\hat{\tau}_{\gate}^{\ora}= n^{-1}\sum_{k=1}^K \sum_{i \in \mik}\omega_{\gate,i}  e_{\gate,i}$. Denote
\begin{align*}
    \mathcal R_n^{\ora}=\dfrac{M}{n^2}\sum_{i=1}^n\sum_{j=1}^n\{\tilde A_{ij}+\mathbbm 1(i=j)\}\left\{\E(\omega_{\gate,i}e_{\gate,i})-\tau_{\gate}\right\}\left\{\E(\omega_{\gate,j}e_{\gate,j})-\tau_{\gate}\right\}. 
\end{align*}

\begin{theorem}
\label{theorem.clt.cre}
Suppose that $K$ is fixed and that Assumptions~\ref{assumption.neighborhoodinterference}, \ref{assumption.conditionalindependence}, and~\ref{assumption.CRE} in the Supplementary Material, together with the $*=\gate$ version of \Cref{assumption.stability}, hold.  Then, $\var(\hat\tau_{\gate}^{\ora})=O(M^{-1})$ and $(\hat\tau_{\gate}^{\adj}-\tau_{\gate})/\sqrt{\var(\hat\tau_{\gate}^{\ora})}\xrightarrow d\mathcal N(0,1)$.
For $\alpha\in(0,1)$, define
$\mathrm{CI}_{\gate}=[\hat\tau_{\gate}^{\adj}\pm
z_{1-\alpha/2}(\hat V_{\gate}^{\adj})_+^{1/2}]$. If $\liminf_{n\to\infty}\mathcal R_n^{\ora}\ge0$, then
$\liminf_{n\to\infty}\pr(\tau_{\gate}\in\mathrm{CI}_{\gate})\ge1-\alpha$, and if
$\mathcal R_n^{\ora}\to0$, then
$\pr(\tau_{\gate}\in\mathrm{CI}_{\gate})\to1-\alpha$.
Under partial interference within the clusters, the nonnegativity condition holds automatically.
\end{theorem}

\section{Simulation}
\label{sec:simulation}

We conduct three simulation studies targeting the main methodological contributions of the paper: high-dimensional linear adjustment with and without cross-fitting, the bias induced by omitting neighborhood exclusion, and the choice between fixed and diverging $K$ in dense networks.
Additional results in the \sm\ examine robustness across network structures, compare first- and second-order neighborhood exclusion for direct-effect estimation, evaluate variance-optimal and CI-length-optimal adjustments, compare cluster-level and unit-level splitting, and assess GATE inference under cluster randomization.

All simulations reported in the main text assume that $Z_i \stackrel{\mathrm{i.i.d.}}{\sim} \mathrm{Bernoulli}(0.5)$.
For each finite population, the network, covariates, and potential outcomes are generated once and then held fixed, with randomness arising solely from repeated treatment randomization, following the design-based perspective.
Unless otherwise stated, each finite population is evaluated over $1{,}000$ independent treatment assignments.
We report empirical bias, standard deviation (SD), mean squared error (MSE) or root mean squared error (RMSE), average estimated standard error (SE), and the coverage probability (CP) of nominal $95\%$ confidence intervals.

\subsection{Performance as the covariate dimension increases}
\label{sec:simulation.covariate}

We evaluate the finite-sample performance of linear adjustment, with and without cross-fitting, as the covariate dimension diverges.
We consider a partial interference setting with $500$ clusters of size two, yielding a total sample size of $n=1{,}000$.
The covariate matrix $\bm X\in\mathbb R^{n\times d}$ has dimension $d=\lfloor n^{\upsilon}\rfloor$, where $\upsilon\in\{0.50,0.55,0.60,0.65,0.70,0.75\}$.
The entries of $\bm X$ are generated independently from a Student's $t$-distribution with $5$ degrees of freedom.

We independently generate $\alpha_i\stackrel{\mathrm{i.i.d.}}{\sim}\mathcal{N}(1,1)$, $\theta_i\stackrel{\mathrm{i.i.d.}}{\sim}t_3(1,0.5)$, and $\gamma_{ij}'\stackrel{\mathrm{i.i.d.}}{\sim}t_3(1,0.5)$, where $t_3(\mu,\sigma)$ denotes a Student's $t$-distribution with $3$ degrees of freedom, location parameter $\mu$, and scale parameter $\sigma$.
To stabilize the overall magnitude of spillover effects, we normalize the spillover coefficients as
$\gamma_{ij}=\gamma_{ij}'\mathbbm{1}(n_i>0)/\max\{n_i,1\}$.
The treatment-dependent component is
$U_i=\alpha_i+\theta_i(Z_i-0.5)+\sum_{j=1}^nA_{ij}\gamma_{ij}(Z_j-0.5)$,
and the outcome is generated as
$Y_i=U_i+\bm X_i^\top\bm\beta$, where $\bm\beta=(1,\ldots,1)^\top/\sqrt{d}$.

\begin{figure}[!htbp]
\centering
\includegraphics[width=0.9\textwidth]{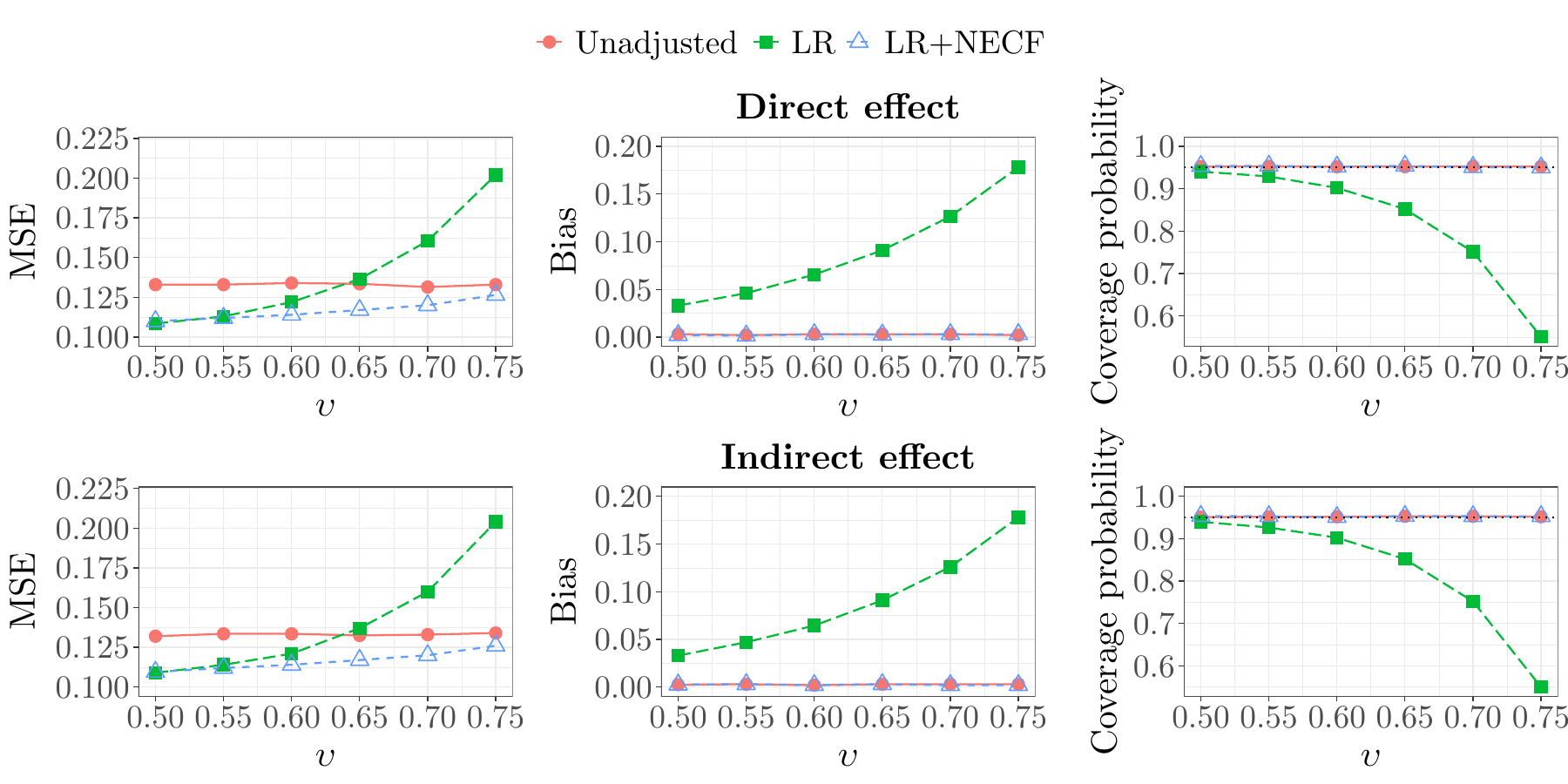}
\caption{Simulation results for linear adjustment as the covariate dimension increases.}
\label{fig:direct.indirect}
\end{figure}

We compare three estimators: (i) the unadjusted estimator; (ii) the linear regression-adjusted estimator based on the CI-length-optimal criterion without sample splitting (``LR''); and (iii) its cross-fitted counterpart (``LR+NECF'').
Under partial interference, sample splitting is performed at the cluster level by randomly partitioning the $500$ clusters into $K=2$ folds.
Across $50$ independently generated finite populations, \Cref{fig:direct.indirect} reports the median bias, MSE, and coverage probability for the direct effect (top panel) and indirect effect (bottom panel).
As the covariate dimension $d=\lfloor n^{\upsilon}\rfloor$ increases, LR deteriorates markedly, exhibiting increasing bias, larger MSE, and coverage well below the nominal $95\%$ level.
In contrast, LR+NECF remains stable across all values of $\upsilon$, with negligible bias, consistently lower MSE, and coverage probabilities close to the nominal level.

\subsection{Impact of omitting neighborhood exclusion}
\label{sec:cf.comparison}

This study examines the bias induced by conventional cross-fitting procedures that construct training samples without excluding neighboring units.
We consider a partial-interference design with $M=500$ disjoint clusters, each containing two units, yielding a total sample size of $n=1{,}000$.
The covariate dimension is set to $d=\lfloor M^{0.75}\rfloor$.
For each cluster, we generate a cluster-level covariate vector $\bm B_m\stackrel{\mathrm{i.i.d.}}{\sim}\mathcal N(\bm 0,\bm I_d)$ and assign the same covariates to both units in the cluster, i.e., $\bm X_i=\bm B_m$ for $i\in\mathcal C_m$.
The coefficient vector is set to $\bm\beta=(1,\ldots,1)^\top/\sqrt d$.
The outcomes are generated as $Y_i=6\bm X_i^\top\bm\beta+6(Z_i-Z_{i'})+\varepsilon_i$, where $i'$ denotes the other unit in the same cluster and $\varepsilon_i\stackrel{\mathrm{i.i.d.}}{\sim}\mathcal N(0,1)$.

We compare three cross-fitting procedures, each using $K=10$ folds:
(i) \textbf{Unit-level splitting without neighborhood exclusion (``Included'').}
The sample is partitioned at the unit level, and the outcome models are fitted without excluding neighbors of the evaluation units from the training sample.
(ii) \textbf{Unit-level splitting with neighborhood exclusion (``Excluded'').}
The sample is partitioned at the unit level, with neighbors of the evaluation units removed from the training sample.
(iii) \textbf{Cluster-level splitting (``Cluster'').}
The sample is partitioned at the cluster level, so that all units within the same cluster are assigned to the same fold.
Under partial interference, this splitting scheme automatically satisfies the neighborhood-exclusion requirement.

\begin{figure}[ht]
\centering
\includegraphics[width=0.9\textwidth]{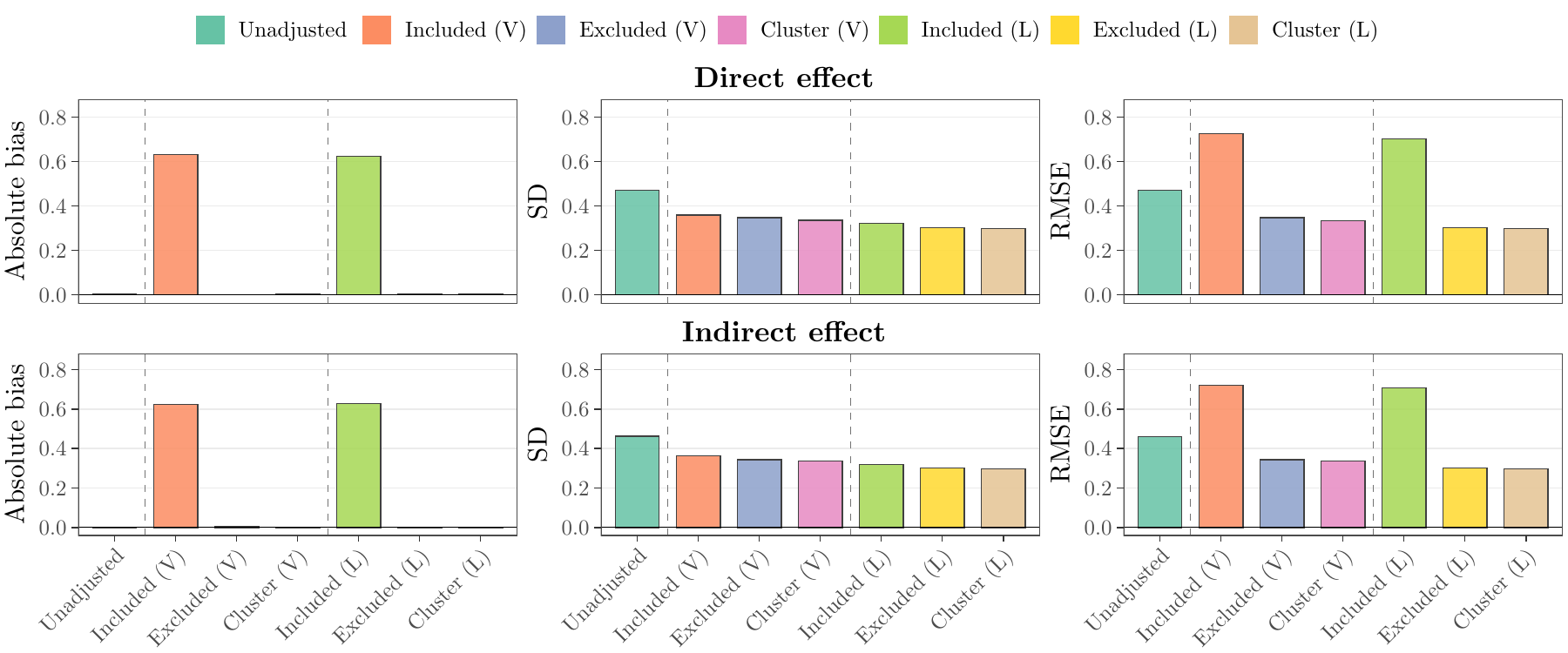}
\caption{Bar plots comparing cross-fitting schemes.}
\label{fig:cf.comparison}
\end{figure}

\Cref{fig:cf.comparison} reports the absolute bias, SD, and RMSE, from left to right, for the direct-effect estimators in the top row and the indirect-effect estimators in the bottom row.
The suffixes ``(V)'' and ``(L)'' denote the variance-optimal and CI-length-optimal adjustments, respectively, while the unadjusted estimator is included as a benchmark.
Three main patterns emerge.
First, and most importantly, unit-level splitting without neighborhood exclusion (``Included'') produces substantial absolute bias for both effects.
Moreover, the ``Included'' estimators generally exhibit larger SDs than their ``Excluded'' and ``Cluster'' counterparts, further degrading their performance.
The combination of bias and increased variability leads to substantially larger RMSEs.
Second, ``Cluster'' generally yields slightly smaller SDs and RMSEs than ``Excluded'', consistent with our theoretical results.
Third, the CI-length-optimal adjustment exhibits smaller finite-sample SDs than the variance-optimal adjustment under all three splitting procedures.
This finite-sample ordering is the reverse of their oracle-variance ordering: although the variance-optimal adjustment has the smaller oracle variance, its additional network-induced cross-unit covariance terms may slow convergence toward the oracle limit.
\Cref{sec.sm:var.vs.CI} of the \sm\ confirms this oracle-variance ordering.

\subsection{Choice of \texorpdfstring{$K$}{K} in dense networks}
\label{sec:diverging.k}

This study investigates how the choice of $K$ affects finite-sample performance as network density increases.
The theoretical concern is that fixed-$K$ cross-fitting may leave too few effective training observations after neighborhood exclusion in dense networks.
We generate networks from Erd\H{o}s--R\'enyi graphs with varying numbers of nodes ($n$) and expected neighborhood sizes ($n\rho_n$).
We generate covariates $\bm X\in\mathbb R^{n\times 2}$ with entries independently drawn from a $\mathrm{Unif}(0,1)$ distribution, and generate outcomes as
$Y_i=\{1+\exp(-U_i)\}^{-1}+X_{i,1}+X_{i,2}$,
where $U_i$ follows the specification in \Cref{sec:simulation.covariate}.
We compare a fixed-$K$ regime, where the number of folds is fixed at $K=2$, with a diverging-$K$ regime, where the number of folds increases with the neighborhood size and scales with the expected degree.

\Cref{tab:K.comparison.linear} reports the performance of the unadjusted estimator and the CI-length-optimal adjustment under the linear specification.
Under fixed-$K$ splitting, neighborhood exclusion leaves too few observations for effective model fitting, whereas the diverging-$K$ regime remains stable and substantially improves precision.
For direct effects, empty or extremely small training sets produce a highly skewed mixture with rare, jointly large estimation errors and estimated standard errors; hence the empirical SD can greatly exceed the average SE even when studentized coverage is near nominal.
For indirect effects, second-order exclusion makes the training sets empty in these configurations; following Algorithm~\ref{alg:necf}, the fitted function is then set to zero, so the fixed-$K$ estimator reduces exactly to the unadjusted estimator, explaining their identical results.
Thus, the near-nominal coverage under fixed $K$ reflects fallback or self-normalization rather than successful adjustment, supporting the use of $K$ growing with neighborhood size.

\begin{table}[ht]
\centering
\renewcommand{\arraystretch}{0.75}
\setlength{\tabcolsep}{4pt}
\begin{threeparttable}
\caption{Comparison of fixed-$K$ and diverging-$K$ splitting}
\label{tab:K.comparison.linear}
\begin{tabular}{ccccccccc}
\toprule
$n$ & $n\rho_n$ & Estimand & Method & Bias & SD & RMSE & SE & CP \\
\midrule 
\multirow{3}{*}{500} & \multirow{3}{*}{10} & \multirow{3}{*}{$\tau_{\dir}=0.177$} & unadj & -0.006 & 0.151 & 0.151 & 0.152 & 0.951 \\
&&& Fixed-$K$ ($K=2$) & 0.007 & 0.590 & 0.589 & 0.277 & 0.952 \\
&&& Diverging-$K$ ($K=20$) & -0.001 & 0.033 & 0.033 & 0.036 & 0.964 \\
\midrule 
\multirow{3}{*}{1000} & \multirow{3}{*}{20} & \multirow{3}{*}{$\tau_{\dir}=0.175$} & unadj & -0.005 & 0.111 & 0.111 & 0.111 & 0.951 \\
&&& Fixed-$K$ ($K=2$) & 0.122 & 3.611 & 3.612 & 0.399 & 0.951\\
&&& Diverging-$K$ ($K=40$) & -0.002 & 0.023 & 0.023 & 0.025 & 0.962 \\
\midrule 
\multirow{3}{*}{500} & \multirow{3}{*}{10} & \multirow{3}{*}{$\tau_{\ind}=0.170$} & unadj & -0.021 & 1.522 & 1.521 & 1.508 & 0.945 \\
&&& Fixed-$K$ ($K=2$) & -0.021 & 1.522 & 1.521 & 1.508 & 0.945 \\
&&& Diverging-$K$ ($K=50$) & -0.019 & 0.523 & 0.523 & 0.525 & 0.951 \\
\midrule 
\multirow{3}{*}{1000} & \multirow{3}{*}{20} & \multirow{3}{*}{$\tau_{\ind}=0.173$} & unadj & -0.102 & 2.216 & 2.217 & 2.177 & 0.941 \\
&&& Fixed-$K$ ($K=2$) & -0.102 & 2.216 & 2.217 & 2.177 & 0.941 \\
&&& Diverging-$K$ ($K=120$) & -0.039 & 0.991 & 0.992 & 0.976 & 0.946 \\
\bottomrule
\end{tabular}
\end{threeparttable}
\end{table}

\section{Real-data analysis}
\label{sec:realdata}

We revisit the anti-conflict intervention experiment conducted during the 2012--2013 school year in 56 public middle schools in New Jersey \citep{paluck2016changing}.
The intervention aimed to reduce hostile behaviors, including bullying and social exclusion, by shifting social norms within schools.
The experiment used a two-stage stratified design: 28 schools were randomly assigned to treatment, and within each treated school, half of the seed-eligible students were randomized to receive an invitation to the anti-conflict program within gender-by-grade blocks.
Following \citet{leung2022causal} and \citet{gao2025causal}, we focus on seed-eligible students in the five largest treated schools.
Although invitations were block randomized by gender and grade, we follow \citet{leung2022causal} and approximate the assignment mechanism by independent Bernoulli randomization with $r=0.5$, matching the marginal treatment probability under the stratified design.

The outcome of interest is a self-reported indicator of wristband wearing, where wristbands were awarded to students who exhibited anti-conflict behavior.
The network is constructed from survey nominations: we set $A_{ij}=1$ if student $i$ nominated student $j$ as someone with whom they had recently interacted.
After excluding observations with missing outcomes, the final analysis sample consists of 265 students, 144 of whom were assigned to treatment.
The resulting network $\bm A$ has density $\rho_n=0.002$, with maximum in-degree 5 and maximum out-degree 3.
The corresponding second-order adjacency matrix $\tilde{\bm A}$ has density $\tilde{\rho}_n=0.004$ and maximum degree 7.

For covariate adjustment, we use 20 covariates: seven binary indicators (gender, grade, race, internet access, Facebook use, younger siblings, and new-student status), three numerical variables (height, number of absences, and GPA), and their corresponding in-neighborhood averages, defined as zero for units with no in-neighbors.
We convert the categorical height variable to inches using representative values of 52, 55, 58, 61, 64, 67, and 70 inches, impute missing height values as 61 inches, and set missing binary indicators to 0.
Because the outcome is binary, we consider both linear adjustment and logistic regression followed by calibration, each under the variance-optimal and CI-length-optimal criteria.
We use 10-fold cluster-level sample splitting, with clusters constructed in advance by applying the Leiden algorithm with a modularity objective to the observed network.
To reduce sensitivity to a particular random split, we also consider a repeated-splitting aggregation procedure and report the corresponding results in \Cref{sec.sm:realdata.repeated} of the \sm.

\Cref{tab:realdata} shows that covariate adjustment shortens the confidence intervals for both the direct and indirect effects relative to the unadjusted estimator.
In the table, ``LR-V'' and ``LR-L'' denote the variance-optimal and CI-length-optimal adjustments, respectively.
Because $n\rho_n=0.555$ and $n\tilde{\rho}_n=1.004$, we treat the network as sparse and use the AS variance estimator for the primary analysis; intervals based on the dense-network variance estimator are reported as exploratory.
Logistic regression with calibration yields the largest reduction in interval length for the indirect effect, whereas linear and logistic adjustments perform similarly for the direct effect.
The AS intervals are longer than the exploratory dense-network intervals, consistent with the theoretical conservativeness of the AS procedure in sparse networks.
Additional sensitivity analyses for the choice of $K$ and the number of covariates are reported in \Cref{sec.sm:realdata.sensitivity} of the \sm.

\begin{table}[ht]
\centering
\renewcommand{\arraystretch}{0.75}
\setlength{\tabcolsep}{4pt}
\begin{threeparttable}
\caption{Results for the anti-conflict intervention.}
\label{tab:realdata}
\begin{tabular}{ccccccc}
\toprule
& & & \multicolumn{2}{c}{AS (primary)}
& \multicolumn{2}{c}{Dense (exploratory)} \\
\cmidrule(lr){4-5}\cmidrule(lr){6-7}
Estimand & Method & Estimate & 95\% CI & Length & 95\% CI & Length \\
\midrule
\multirow{5}{*}{$\tau_{\dir}$}
& unadj & 0.294 & (0.083, 0.506) & 0.423 & (0.130, 0.458) & 0.328 \\
& LR-V & 0.204 & (0.029, 0.379) & 0.350 & (0.070, 0.338) & 0.268 \\
& LR-L & 0.210 & (0.046, 0.374) & 0.328 & (0.084, 0.336) & 0.252 \\
& logit+LR-V & 0.179 & (0.015, 0.343) & 0.328 & (0.052, 0.305) & 0.253 \\
& logit+LR-L & 0.190 & (0.028, 0.352) & 0.324 & (0.065, 0.315) & 0.250 \\
\midrule
\multirow{5}{*}{$\tau_{\ind}$}
& unadj & 0.106 & (-0.101, 0.313) & 0.414 & (-0.052, 0.263) & 0.314 \\
& LR-V & 0.051 & (-0.139, 0.240) & 0.379 & (-0.077, 0.178) & 0.255 \\
& LR-L & 0.094 & (-0.073, 0.262) & 0.334 & (-0.010, 0.199) & 0.209 \\
& logit+LR-V & 0.077 & (-0.074, 0.228) & 0.302 & (-0.021, 0.174) & 0.195 \\
& logit+LR-L & 0.078 & (-0.069, 0.224) & 0.293 & (-0.016, 0.172) & 0.187 \\
\bottomrule
\end{tabular}
\end{threeparttable}
\end{table}

\vspace{-2em}

\section{Discussion}
\label{sec:discussion}

This paper develops neighborhood-excluded cross-fitting for design-based covariate adjustment under network interference.
By separating the treatment assignments affecting the training outcomes from those entering the evaluation Horvitz--Thompson weights, the procedure restores the conditional independence that ensures finite-sample unbiasedness.
We establish asymptotically valid Wald inference for direct and indirect effects under Bernoulli randomization and for the GATE under Bernoulli cluster randomization.
For high-dimensional linear adjustment, we derive variance-optimal and CI-length-optimal adjustments and explicit rate conditions allowing the covariate dimension to diverge. The framework can also incorporate fitted values from flexible prediction methods as additional covariates through calibration.

Two directions merit further investigation.
First, extending the framework to other randomized designs, such as complete randomization or rerandomization, requires sample-splitting and exclusion rules that account for dependence induced by the assignment mechanism.
Second, while our theory characterizes how the number of folds should scale with neighborhood size, network-aware splitting strategies, such as community detection or graph partitioning, may retain more training data and improve predictive accuracy, especially in dense networks, while preserving the required conditional independence.


\bibliography{causal}

\end{document}